\documentclass[11pt]{article}

\usepackage[margin=1in]{geometry}
\usepackage{amsmath,amssymb,amsfonts,bm}
\usepackage{booktabs}
\usepackage{multirow}
\usepackage{graphicx}
\usepackage{float}
\usepackage{array}
\usepackage{enumitem}
\usepackage[numbers,sort&compress]{natbib}
\usepackage[colorlinks=true,linkcolor=blue,citecolor=blue,urlcolor=blue]{hyperref}
\usepackage{xurl}
\usepackage{setspace}
\usepackage{authblk}
\usepackage{caption}
\usepackage{tikz}
\usetikzlibrary{arrows.meta,positioning,calc,fit}
\usepackage{xcolor}
\definecolor{commentcolor}{RGB}{200, 30, 30}

\definecolor{suggestcolor}{RGB}{30, 80, 200}

\usepackage[normalem]{ulem}

\definecolor{colorouns}{RGB}{200, 30, 200}

\definecolor{colorjon}{RGB}{0, 0, 200}

\definecolor{fjgreen}{RGB}{0,100,0}

\newcommand{\fjsout}{\bgroup\markoverwith{\textcolor{fjgreen}{\rule[0.5ex]{2pt}{0.6pt}}}\ULon}

\usepackage{cancel}

\definecolor{colorQ}{RGB}{0,155,175}

\definecolor{colorRev}{RGB}{128,0,160}

\definecolor{colorFlag}{RGB}{230,115,0}

\title{\textbf{Screening strongly correlated materials in solid oxide cells with quantum simulation}}

\author[1]{Ouns El Harzli}
\author[2]{Jelle Bor}
\author[2,3]{Linde Wester}
\author[1]{Leonid Zhukov}
\author[1]{Hanl Park}
\author[1]{Jean-François Bobier}
\author[1]{Matt Langione}
\author[4,5,6,7]{Antonino Marco Saitta}
\author[8]{Jonathan Owens}
\author[8]{Richard Hart}
\author[1,9, \ensuremath{*}, \ensuremath{\dagger}]{\authorcr Yudong Cao}
\author[2, \ensuremath{*}, \ensuremath{\dagger}]{François Jamet}

\affil[1]{BCG X AI Science Institute}
\affil[2]{Alice \& Bob}
\affil[3]{EVE Quantum Labs}
\affil[4]{Laboratoire de Physique de l'ENS, Ecole Normale Supérieure PSL}
\affil[5]{Sorbonne Université}
\affil[6]{CNRS}
\affil[7]{Institut Universitaire de France}
\affil[8]{GE Vernova Advanced Research}
\affil[9]{Zapata Quantum}

\date{}
\begin{document}

\maketitle
\begingroup
\renewcommand{\thefootnote}{\fnsymbol{footnote}}

\footnotetext[1]{%
  Corresponding authors:
  \texttt{ycao@zapataquantum.com};
  \texttt{francois.jamet@alice-bob.com}.
}

\footnotetext[2]{%
  These authors contributed equally to the
  supervision of this work.
}

\endgroup

\begin{abstract}
Solid oxide fuel cells (SOFCs) and solid oxide electrolysis cells (SOECs) offer high efficiency, fuel flexibility, and their high operating temperature enables direct thermal integration with fuel reforming and industrial waste-heat streams, but their commercialization is constrained by the performance--durability trade-off of oxygen electrodes, especially at intermediate temperatures. 
Perovskite and related mixed ionic-electronic conducting oxides are attractive oxygen-electrode materials because they combine electronic conductivity, oxygen-ion conductivity (enabled by their oxygen non-stoichiometry), and catalytic activity.
Yet, their predictive design remains difficult because of strongly correlated transition-metal \(d\) states. 
This challenge is amplified when interfacial chemistry with the electrolyte materially changes vacancy energetics, reaction rates, and degradation pathways.
While specific methods are still in development, quantum computers are expected to be well suited for simulating strongly correlated systems and could therefore help overcome this bottleneck in predictive design.

In this paper, we develop an end-to-end framework for assessing a quantum-computing application in industrial materials science: the screening of strongly correlated oxygen-electrode materials for SOFCs and SOECs.
First, we formulate the materials problem in terms of interface observables: oxygen-vacancy formation energies, vacancy migration barriers, and electronic conductivity.
Second, we describe how these observables can be obtained using a diagrammatic GW-based embedding, either GW+DMFT or self-energy embedding theory (SEET), applied to a DFT-screened shortlist.
The atomistic electrode/electrolyte interface defines the electronic-structure problem from which the correlated impurity models are constructed.
Quantum impurity solvers would provide the finite-temperature impurity Green's functions required by the embedding \citep{BauerEtAl2016}.
The computational challenges that classical solvers may encounter as the interacting active space grows, together with the polynomial scaling of quantum Hamiltonian simulation for efficiently represented discretized models, open an opportunity for quantum advantage.
We identify the benchmarks needed to assess this opportunity through convergence of vacancy and migration energetics and comparison with the best applicable classical solvers.
We provide illustrative logical-qubit counts and identify the thermal-state preparation, simulation, measurement, and repeated-solve costs needed for a complete resource assessment.
Third, we connect microscopic observables to macroscopic engineering metrics using a continuum model for mixed ionic--electronic conducting oxygen electrodes that couples oxygen surface exchange, solid-state oxygen transport, electrode microstructure, and finite electrode thickness.
Finally, we estimate the potential worldwide business value of improved electrode materials under SOFC and SOEC deployment scenarios informed by public technical targets and electrolyser deployment data.

Our central contribution is an end-to-end description of the quantum computing application, produced by a chain of logic that flows from an industrial materials-screening problem to quantum-amenable impurity models, then onward to cell-level performance metrics and economic value. 
Due to the wide applicability of DMFT-based embedding for strongly correlated material simulations, our approach could be replicated to a range of material design use cases of similar potential economic value.
This provides a tangible example of where and how quantum computers could play a role in future material design workflows.
\end{abstract}

{
\hypersetup{linkcolor=black}
\tableofcontents
}

\section{Introduction}
\begin{figure*}[t!]
    \centering
    \includegraphics[width=8cm]{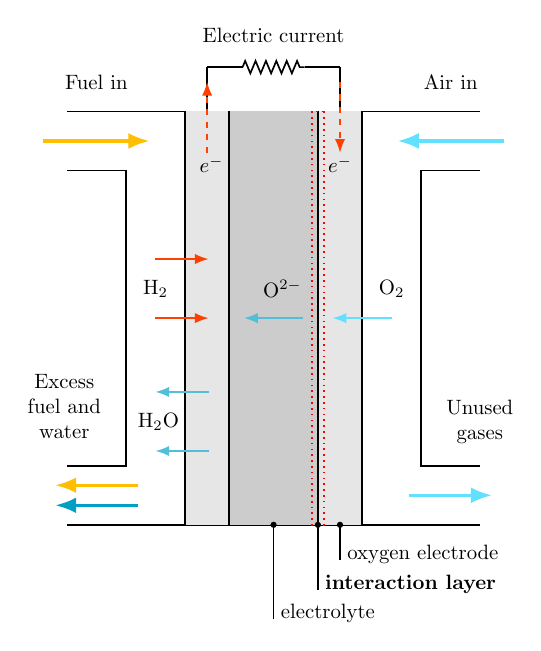}
    \caption{Schematic of a SOFC running on air and a fuel that contains H\(_2\). 
    At the cathode, \(\mathrm{O_2 + 4e^- \rightarrow 2O^{2-}}\); the \(\mathrm{O^{2-}}\) ions cross the solid ceramic electrolyte to the anode, where \(\mathrm{H_2 + O^{2-} \rightarrow H_2O + 2e^-}\). 
    Since the electrolyte blocks electrons, they must flow through the external circuit, generating current; net reaction: \(\mathrm{2H_2 + O_2 \rightarrow 2H_2O}\). 
    Running it in reverse, i.e.~applying an external voltage instead of drawing current, the same cell operates as a SOEC, splitting H\(_2\)O back into H\(_2\) and O\(_2\).}
    \label{fig:SOFS}
\end{figure*}

Solid oxide cells are among the most compelling electrochemical energy-conversion technologies because they can operate reversibly, reach high electrical efficiency, and integrate naturally with process heat, fuels, and industrial systems \citep{He2023CriticalReview,Yang2024ReversibleSOCs}. 
In SOFC mode they convert chemical fuels into electricity with system efficiencies exceeding 60\%, while in SOEC mode they can achieve high-temperature electrolysis (i.e.\ converting electricity in chemical fuels)  with electricity consumption targets materially below low-temperature electrolysis \citep{DOE_SOFC_Program,DOE_HTE_Targets}. 
The operating principle of the SOFC, including the electrode half-reactions and electron flow through the external circuit, is illustrated in Fig.~\ref{fig:SOFS}.
SOFCs are primarily used for stationary electricity generation, including distributed power for data centers, commercial and industrial facilities, microgrids, and combined heat-and-power systems, whereas SOECs are primarily used to convert electricity and steam into hydrogen or synthetic-fuel feedstocks and thereby supporting renewable-energy storage and industrial decarbonization \citep{DOE2020SOFCApplications,DOE2024SolidOxide}. 
The central materials challenge is the oxygen electrode: the air-side electrode in SOFC operation and the oxygen-evolution electrode in SOEC operation.
Lowering operating temperature from the traditional \(800\text{--}1000^\circ\mathrm{C}\) range toward \(500\text{--}700^\circ\mathrm{C}\) reduces balance-of-plant complexity and degradation, but it also increases polarization resistance and exposes difficult trade-offs among vacancy chemistry, transport, electronic conductivity, interface reactions, and thermo-mechanical compatibility \citep{Takahashi2010RuddlesdenPopper,He2023CriticalReview,NisarEtAl2024}.

Perovskite and related mixed ionic--electronic conducting oxides are natural candidates for oxygen electrodes because they can provide large oxygen-vacancy concentrations, fast surface exchange, and high electronic conductivity \citep{He2023CriticalReview,Yang2024ReversibleSOCs}. 
However, the most relevant chemistries contain late transition metals such as Co, Fe, Mn, and Ni, whose local spin states and metal--oxygen covalency can change sharply with composition, oxygen stoichiometry, and local environment. 
More specifically, the computational challenge arising with strongly correlated materials is due to partially filled and spatially localized \(d\) or \(f\) shells that generate local Coulomb interactions that are comparable to the relevant one-electron energy scales.
Electrons therefore cannot always be represented accurately as approximately independent particles: multiple local charge, orbital, and spin configurations can contribute at similar energies, and their relative importance can change with temperature, composition, oxidation state, strain, or defects. 
This physics occurs prominently in transition-metal compounds containing, for example Co, Fe, Mn, or Ni, \(3d\) states and in rare-earth and actinide materials containing localized \(4f\) or \(5f\) states \cite{KotliarEtAl2006,Held2007ElectronicStructure}. 
These are precisely the regimes in which semilocal density functional theory (DFT) and even DFT+\(U\) (i.e.~with a static Hubbard correction) can become unreliable for total energies, charge states, and spectral quantities \citep{KotliarEtAl2006,Held2007ElectronicStructure,IsaacsMarianetti2020}. 
The problem is more severe at interfaces, because the electrolyte changes the local crystal field, hybridization, charge redistribution, and possible reaction pathways. 
In practice, a material that looks promising in bulk may fail once interfaced with electrolytes like yttria-stabilized zirconia (YSZ), gadolinium-doped ceria (GDC), samaria-doped ceria (SDC), or lanthanum strontium gallate magnesite (LSGM) due to altered vacancy energetics, reaction kinetics, interdiffusion, and thermal-expansion mismatch \citep{Yang2004ReactionKineticsLSMYSZ,He2023CriticalReview,NisarEtAl2024}.

Going beyond DFT is necessary to describe the local, dynamical correlations of transition-metal \(d\) electrons. the most widely used method combined DFT with Dynamical Mean field Theory (DMFT) captures these correlations, but its double-counting correction, the subtraction of interactions already included in DFT, is not uniquely defined. This choice can affect vacancy formation and migration energies, so a more accurate impurity solver alone may not improve the ranking of electrode candidates. GW+DMFT \citep{Biermann2003GWDMFT} and self-energy embedding theory (SEET) \citep{Kananenka2015SEET} instead define the overlap between their weakly and strongly correlated treatments diagrammatically. We propose to compare these approaches in a bulk pilot (Section~\ref{sec:tiers}). The screening workflow uses DFT to obtain geometries and assess all candidates, then applies GW-based embedding to a shortlist. Whether this changes candidate rankings remains an open question (Section~\ref{sec:amplification}).

The possible role of a quantum computer is to solve the correlated impurity problems within this embedding. DMFT maps the lattice problem onto self-consistent Anderson impurity models (AIMs) \citep{KotliarEtAl2006,Held2007ElectronicStructure}; Bauer et al.\
identified impurity solving as a natural quantum subroutine \citep{BauerEtAl2016}. Realistic impurity models can involve off-diagonal hybridization, a full Coulomb interaction tensor and, in GW+DMFT, retarded interactions. These features can increase the cost of classical solvers, particularly as the interacting active space grows, opening an opportunity for quantum advantage (Section~\ref{sec:quantum-algos}). Assessing this opportunity requires comparison with the best applicable classical solvers at the accuracy needed for the target observables. The classical GW calculation remains  part of the total workload. Quantum resource requirements, including logical qubits, gate costs, and repeated impurity solves, are discussed in Sections~\ref{sec:qubits} and \ref{sec:open}. The proposed quantum workflow targets fault-tolerant hardware.

\begin{figure*}[t!]
    \centering
    \includegraphics[width=\textwidth]{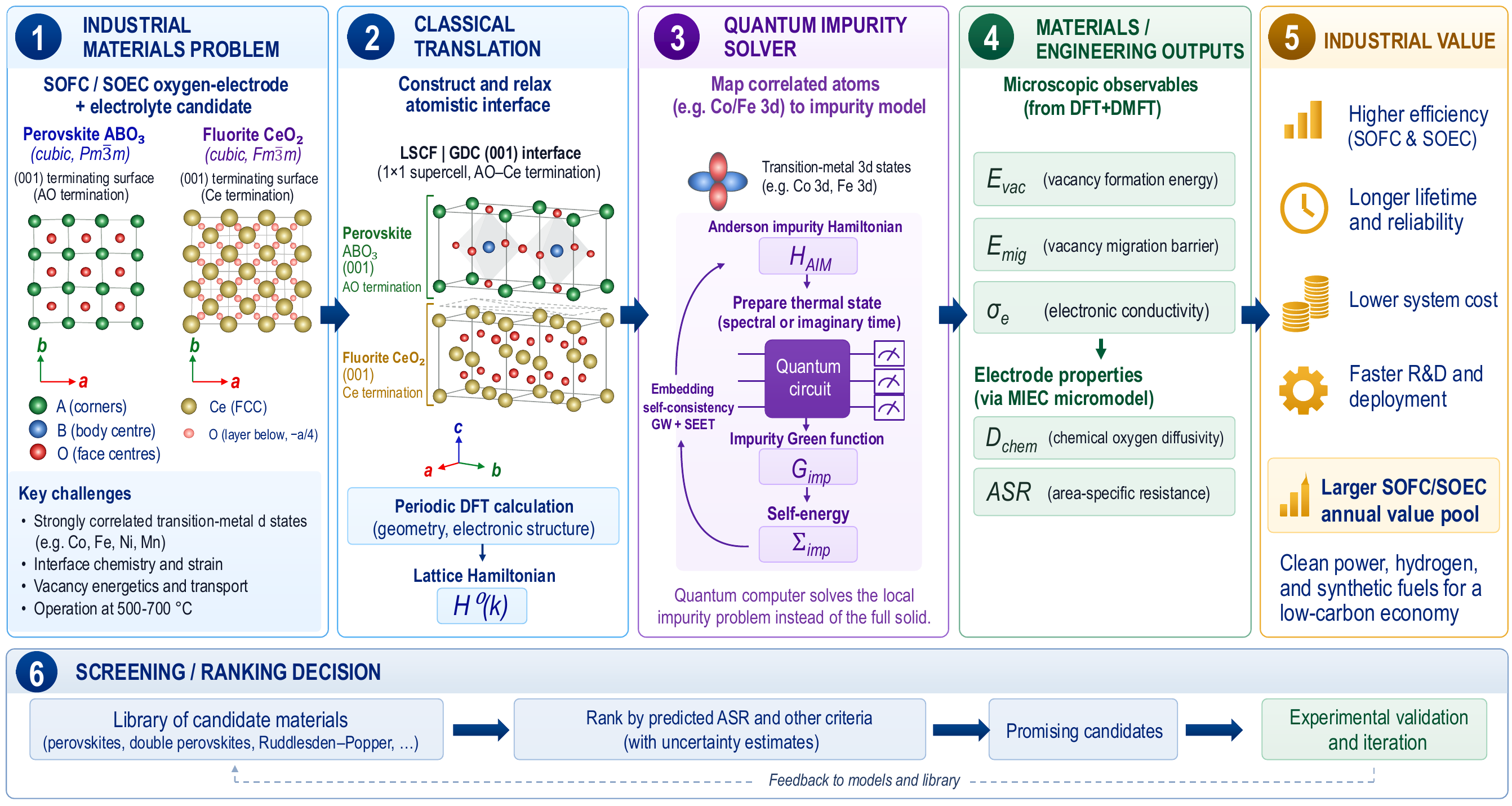}
    \caption{
    End-to-end workflow for quantum-enabled screening of strongly correlated oxygen-electrode materials. 
    Candidate electrode/electrolyte interfaces are first constructed and treated with periodic DFT for all candidates (Tier T0). For the shortlist, a GW-based embedding (GW+DMFT or SEET) maps the strongly correlated local degrees of freedom onto Anderson impurity models (Tier T1), which constitute the quantum-computing subproblem. 
    The resulting interacting Green's functions and self-energies are embedded back into the lattice to calculate microscopic materials observables, including oxygen-vacancy formation energies, migration barriers, and electronic conductivity. 
    These observables are propagated through an electrode microstructure model to engineering quantities such as area-specific resistance and durability, enabling candidate ranking and, ultimately, estimation of industrial value. 
    Experimental validation, performed selectively on the highest-ranked candidates, can further refine the screening models, reducing reliance on the broader, more expensive experimental search.}
    \label{fig:end_to_end_workflow}
\end{figure*}
The gap is not only algorithmic. A useful assessment must connect an industrial materials question to a screening workflow, identify the part a quantum computer could solve, and translate the calculated properties into engineering and business outcomes. This paper develops that connection for solid oxide cell oxygen electrodes, as illustrated in Fig.~\ref{fig:end_to_end_workflow}. Our contributions are fourfold:

\begin{enumerate}[leftmargin=1.5em]
    \item We define an oxygen-electrode screening problem using physically meaningful, interface-aware observables that are relevant to industrial decisions.
    \item We outline a screening workflow in which DFT assesses candidate materials and supplies geometries, while GW-based embedding evaluates a shortlist. We propose a bulk benchmark to compare GW+DMFT and SEET in a bulk pilot, identify the impurity problem that a quantum computer could solve, and discuss the additional work needed to extend the calculation to electrode/electrolyte interfaces.
    \item We connect the calculated materials properties to an electrode microstructure model for area-specific resistance (ASR).
    \item We estimate the worldwide business value of potential materials improvements in both SOEC and SOFC applications.
\end{enumerate}

\section{Related Work}
\subsection{Materials science of oxygen electrodes in solid oxide cells}
The oxygen electrode remains the dominant loss and degradation locus in many SOFC and SOEC architectures, especially under intermediate-temperature operation.
In solid oxide cells, oxygen-electrode materials must combine adequate electronic conductivity, oxide-ion transport, catalytic activity for oxygen reduction/evolution, chemical compatibility with the electrolyte, and thermal-mechanical compatibility \citep{He2023CriticalReview,NisarEtAl2024}. Lanthanum strontium cobalt ferrite (LSCF) and related perovskites exemplify the classical trade-off: they benefit from high oxygen-vacancy concentration and good conductivity, but can suffer from high thermal expansion, cation segregation, and interface instability depending on electrolyte choice \citep{He2023CriticalReview,Yang2024ReversibleSOCs}. 
Interfacial nanocomposites and compositionally tuned interfaces have been shown to materially improve long-term stability, underscoring that the relevant screening problem is not purely about bulk materials \citep{Hu2022InterfacialNanocomposite}. 

\subsection{DFT+DMFT for correlated materials and total energies}
DFT+DMFT is a mature first-principles framework for strongly correlated materials. 
The core idea is to combine a one-electron (mean-field) Hamiltonian from DFT with a many-body treatment of local interactions in a correlated subspace, usually defined by DFT-derived localized orbitals or Wannier functions \citep{KotliarEtAl2006,Held2007ElectronicStructure}. 
Importantly for this work, DFT+DMFT is not only a spectral method; charge-self-consistent formalisms also enable calculation of total energies and hence defect energies, reaction energetics, and migration barriers \citep{ParkEtAl2014}. 
The relevance of these methods to energy materials is underscored by DFT+DMFT studies of battery cathodes such as \(\mathrm{Li}_x\mathrm{CoO}_2\), where phase stability and voltages can differ qualitatively from DFT+\(U\) predictions \citep{IsaacsMarianetti2020}.
For total-energy differences, the main remaining source of uncertainty is the DC correction, which is not uniquely defined: standard prescriptions shift the relative position of the transition-metal \(d\) and oxygen \(p\) states, and with it the energetics \citep{Dang2014Covalency,Haule2015}.
A DC derived from the functional itself has been proposed \citep{Haule2015}; its application to defect energetics in oxides is, to our knowledge, still open.

\subsection{Beyond DFT+DMFT: diagrammatic embedding}
\label{sec:beyond-dmft}
Diagrammatic embeddings start from many-body perturbation theory instead of DFT.
In GW+DMFT, the non-local GW self-energy is combined with the local DMFT self-energy, and the DC term is the local part of the GW self-energy, which is defined diagrammatically and has no free parameter \citep{Biermann2003GWDMFT}; it does, however, depend on the choice of local orbitals and projectors and on the level of self-consistency of the GW step, so the ambiguity is reduced, not eliminated.
SEET follows the same logic, with a weak-coupling treatment of the whole system and a strong-coupling solution of one or more correlated active spaces; its results are systematically improvable as the active space grows \citep{Kananenka2015SEET}, and it has been applied to periodic solids \citep{Rusakov2019SEETPeriodic,Iskakov2020SEETNiO}.

The two options differ where the quantum solver enters.
In GW+DMFT, the impurity interaction is the frequency-dependent screened interaction $U(\omega)$, and the double counting removes both the local GW self-energy and the local polarization \citep{Biermann2003GWDMFT}.
A Hamiltonian-based impurity solver must then represent $U(\omega)$ with auxiliary bosonic modes coupled to the bare high-frequency interaction (Section~\ref{sec:interface_to_impurity_summary}), or reduce it to a static interaction with a renormalized bandwidth \citep{Casula2012Bandwidth}, which reintroduces an approximation.
In SEET, the active space carries the bare static interaction, and the double counting removes the GW self-energy within the active space \citep{Kananenka2015SEET}.
The two also improve along different routes.
SEET enlarges the active space.
GW+DMFT enlarges the impurity to several sites or a cluster.
GW+DMFT is the more widely used of the two, mainly for spectral properties.
We therefore treat both as options for Tier T1 and select between them with a bulk pilot (Section~\ref{sec:tiers}).

To our knowledge, total energies for large relaxed defect supercells of transition-metal oxides have not yet been demonstrated with these methods.
Their classical cost has two parts: the GW step, and the impurity solver, whose requirements and limits are discussed in Section~\ref{sec:quantum-algos}.

\subsection{Classical impurity solver}
\label{sec:quantum-algos}
The quantum opportunity depends on both the size and the structure of the interacting active space. Realistic impurity models can involve off-diagonal hybridization \(\Delta\), a full Coulomb interaction tensor, and retarded interactions, each of which can increase the computational demands on classical solvers. The two Tier~T1 options encounter different combinations of these features. SEET combines enlarged active spaces with the full bare Coulomb tensor and generally off-diagonal hybridization, which can cause a severe, basis-dependent sign problem in hybridization-expansion continuous-time quantum Monte Carlo (CT-HYB) \citep{Shinaoka2015SignProblem}. GW+DMFT introduces a frequency-dependent screened interaction, whose treatment depends on the retained screening channels.

For the monopole screening model of Eq.~\eqref{eq:hbos}, the retarded coupling acts through the total impurity charge \(\hat N_I\), which commutes with the local Hamiltonian even when static spin-flip and pair-hopping terms are retained. This permits a classical treatment of the retarded charge coupling, although the practical cost still depends on the size and structure of the remaining impurity problem. Calculations with density-density interactions have been demonstrated up to five orbitals \citep{Werner2012Satellites}. More general retarded interactions, including dynamical exchange, require different treatments. Ref.~\citep{PhysRevB.92.115123} presents a two-orbital treatment in which spin-flip and pair-hopping terms are approximated as static.

Inchworm QMC provides an alternative for general interactions and hybridizations \citep{Eidelstein2020Inchworm,Strand2024InchwormQMC}; the cited benchmarks concern two-orbital models, and the method retains exponential scaling with the number of interacting orbitals \citep{Eidelstein2020Inchworm}. Exact diagonalization is limited by the combined impurity and bath Hilbert space, while truncated configuration-interaction and tensor-network methods require convergence checks for their respective approximations. Their practical performance for the target models at \(\beta\approx12\)~eV\(^{-1}\) must therefore be established through comparison at the accuracy required for vacancy and migration energetics (Section~\ref{sec:tiers}). The computational challenges that arise as the required interacting active space grows open an opportunity for quantum advantage, to be assessed against the best applicable classical solvers using the total cost of obtaining converged target observables.

\subsection{Electrode micro-modelling and technoeconomics}
At the engineering layer, the appropriate electrode microstructure model depends on the electrode architecture. 
Two-phase composite electrodes in which electronic and ionic transport occur predominantly through distinct phases, for example Ni--YSZ fuel electrodes or conventional LSM--YSZ oxygen electrodes, are naturally described by porous composite-electrode models such as the Costamagna framework \cite{Costamagna1998MicroModelling}. 
By contrast, perovskites such as LSCF are mixed ionic--electronic conductors (MIECs): oxygen incorporation can occur over the internal gas/MIEC surface and oxygen can subsequently diffuse through the MIEC bulk. 
The Adler--Lane--Steele (ALS) continuum framework was developed specifically for this class of porous MIEC oxygen electrode \cite{AdlerLaneSteele1996}.

The ALS framework has subsequently been extended to composite MIEC/ionic-conductor electrodes and to increasingly detailed descriptions of electrode microstructure \cite{MortensenEtAl2014,GoelEtAl2021}. 
A recent review by Panisset et al.\ places these models within the contemporary hierarchy of solid-oxide-cell oxygen-electrode modelling, ranging from homogenized continuum descriptions to reconstructed three-dimensional numerical models \cite{PanissetEtAl2023}. 
For the LSCF/GDC interface running example considered in the rest of the paper, we therefore use a finite-thickness MIEC continuum model within the ALS framework as the screening-level bridge between microscopic materials descriptors and electrode performance, while treating microstructure-resolved finite-element models as a higher-fidelity extension \cite{PanissetEtAl2024}.

On the market side, public technoeconomic targets from Department of Energy (DOE) provide cost, efficiency, and durability reference points for SOFC and high-temperature electrolysis, while International Energy Agency (IEA) data provides the broader deployment context for electrolyser capacity growth \citep{DOE_SOFC_Program,DOE_HTE_Targets,IEA_Electrolysers_2024,IEA_GlobalHydrogenReview2024}. 
Our work combines these literatures into one end-to-end application architecture.

\section{Materials-science formulation of the industrial problem}
\subsection{Use case and screening objective}
We consider oxygen electrodes for SOFC and SOEC systems, with a focus on perovskite and related MIEC such as
\[
\mathrm{La}_{1-x}\mathrm{Sr}_{x}\mathrm{CoO}_3,\quad
\mathrm{La}_{1-x}\mathrm{Sr}_{x}\mathrm{FeO}_3,\quad
\mathrm{La}_{1-x}\mathrm{Sr}_{x}\mathrm{Co}_{1-y}\mathrm{Fe}_{y}\mathrm{O}_3,
\]
where \(x\) and \(y\) denote the fractions of Sr substituted for La and Fe substituted for Co, respectively, and adjacent chemistries that may be preferable for durability, chromium tolerance, or thermal expansion.
The industrial objective is to identify electrode/electrolyte pairs that maintain acceptable polarization resistance and durability at lower operating temperatures, ideally in the \(500\text{--}700^\circ\mathrm{C}\) range for intermediate-temperature devices \citep{Takahashi2010RuddlesdenPopper,NisarEtAl2024,He2023CriticalReview}.

\subsection{Interface observables}
We consider three target observables that capture complementary aspects of oxygen-electrode performance.
Oxygen vacancy formation energy $E_{\mathrm{vac}}$ measures how energetically favorable it is to create an empty oxygen site, which determines the equilibrium vacancy concentration available for oxygen-ion transport.
Oxygen vacancy migration energy $E_{\mathrm{mig}}$ measures the barrier for an oxygen ion to move into a neighboring vacancy, and therefore controls the ionic conductivity of the electrode and its contribution to the overall polarization resistance. 
Electronic conductivity $\sigma_e$ measures how efficiently electrons can move through the electrode to the reaction sites, enabling charge transfer during the oxygen reduction or evolution reaction.
Together, they characterize vacancy availability, oxygen-ion mobility, and electronic transport: three key ingredients of a high-performance SOFC/SOEC oxygen electrode, for which low $E_{\mathrm{vac}}$ and $E_{\mathrm{mig}}$ (favoring high vacancy concentration and fast ionic transport) together with high $\sigma_e$ (favoring low electronic resistance) are generally desirable. 
These observables in an interface can differ significantly from their bulk values, and so must be computed for the electrode/electrolyte interface, since interfacial effects can shift vacancy energetics, migration barriers, and conductivity from the bulk:
\[
\left(
E_{\mathrm{vac}}^{\mathcal I},
E_{\mathrm{mig}}^{\mathcal I},
\sigma_e^{\mathcal I}
\right),
\]
where \(\mathcal I\) denotes the dependence in the interface. 

The interfacial oxygen-vacancy formation energy is
\begin{equation}
    E_{\mathrm{vac}}^{\mathcal I} = E_{\mathrm{tot}}^{V_{\mathrm O}} + \mu_{\mathrm O}(T,p_{\mathrm O_2}) - E_{\mathrm{tot}},
\label{eq:evac}
\end{equation}
where \(V_{\mathrm O}\) denotes an oxygen vacancy; \(E_{\mathrm{tot}}\) and \(E_{\mathrm{tot}}^{V_{\mathrm O}}\) are the total energies of the defect-free and vacancy-containing interface structures, respectively; and \(\mu_{\mathrm O}(T,p_{\mathrm O_2})\) is the chemical potential of the removed oxygen atom at temperature \(T\) and oxygen partial pressure \(p_{\mathrm O_2}\), obtained from DFT calculation of an isolated oxygen molecule, supplemented by experimental thermochemical corrections \cite{Chase1998JANAF}.

The reference for \(\mu_{\mathrm O}\) must be consistent with the level of theory used for \(E_{\mathrm{tot}}\) and \(E_{\mathrm{tot}}^{V_{\mathrm O}}\).
If the solid is treated with a correlated embedding while the O\(_2\) reference comes from semilocal DFT, errors of the molecular reference, a known weakness of semilocal functionals \citep{WangMaxischCeder2006}, do not cancel in Eq.~\eqref{eq:evac} and enter every vacancy energy as a common shift.
The oxygen reference must therefore be computed at the same level as the solid, or calibrated against a measured oxidation enthalpy of a reference oxide, before absolute values of \(E_{\mathrm{vac}}\) are compared with experiment.
Energy differences between candidates, which are what the screen needs, are less sensitive to this shift.

The interfacial migration barrier is given by the energy difference between the transition state and the initial state along the migration pathway of an oxygen ion into a neighboring vacancy:
\begin{equation}
    E_{\mathrm{mig}}^{\mathcal I} = E_{\mathrm{TS}} - E_{\mathrm{IS}},
\label{eq:emig}
\end{equation}
where \(E_{\mathrm{IS}}\) is the total energy of the interface structure with the oxygen ion in its initial lattice site, and \(E_{\mathrm{TS}}\) is the total energy at the saddle point of the migration pathway, i.e.~the highest-energy configuration the ion passes en route to the neighboring vacancy. 

For electronic transport we target the interface conductivity given by:
\begin{equation}
    \sigma_e^\mathcal{I} = \frac{J}{E}.
    \label{eq:sigmae}
\end{equation}
where $J^\mathcal{I}$ is the electronic current density through the interface and $E^\mathcal{I}$ is the applied electric field at the interface. In practice, \(\sigma_e^{\mathcal I}\) is evaluated from the Green's functions via a Kubo-formula expression, rather than from an explicitly applied field.

These observables serve complementary roles. Together, \(E_{\mathrm{vac}}^{\mathcal I}\), \(E_{\mathrm{mig}}^{\mathcal I}\), and \(\sigma_e^{\mathcal I}\) determine transport and utilization. 
Computing these observables at the interface is important because a material can exhibit favorable vacancy formation, easy ion migration, and high conductivity as a bulk material, yet still fail through electrolyte reactivity or interface degradation \citep{Yang2004ReactionKineticsLSMYSZ,He2023CriticalReview}. 

\subsubsection[Computing interface observables via embedding and quantum-computable quantities]{Computing interface observables via embedding and quantum-computable quantities}
\label{sec:observables}
\paragraph{Total energies.}
The total energies entering the expressions for \(E_{\mathrm{vac}}^{\mathcal I}\) and \(E_{\mathrm{mig}}^{\mathcal I}\) are computed with the embedding as follows. We first write the familiar DFT+DMFT form, which serves as the reference, and then state what changes in Tier~T1.
Although these quantities describe oxygen-vacancy formation and ion migration, both are obtained from differences of electronic total energies evaluated for different atomic configurations.
Because DMFT recomputes the one-electron energy using the interacting Green's function, the corresponding DFT band-energy contribution must first be removed.
The contribution already represented by DFT within the correlated orbitals must also be subtracted before adding the explicit local interaction energy, to avoid double counting.
For a given atomic configuration \(X\),
\begin{equation}
    E_{\mathrm{tot}}^{X}
    =
    E_{\mathrm{DFT}}
    -
    E_{\mathrm{band,DFT}}
    +
    \mathrm{Tr}\!\left[H_{\mathrm{KS}}G_{\mathrm{latt}}\right]
    +
    \left\langle H_{\mathrm{int}}\right\rangle
    -
    E_{\mathrm{dc}}.
    \label{eq:etot}
\end{equation}
Here, \(E_{\mathrm{DFT}}\) is the DFT total energy, \(E_{\mathrm{band,DFT}}\) removes the corresponding band-energy contribution, \(H_{\mathrm{KS}}\) is the Kohn--Sham Hamiltonian, and \(G_{\mathrm{latt}}\) is the interacting lattice Green's function.
The trace includes the frequency sum and orbital, spin and wavevector indices.
The terms \(\langle H_{\mathrm{int}}\rangle\) and \(E_{\mathrm{dc}}\) account, respectively, for the explicit local interaction energy and the double-counting correction.
All quantities are evaluated for configuration \(X\), with consistent charge-density bookkeeping.

Equation~\eqref{eq:etot} is the DFT+DMFT form.
In Tier~T1, GW+DMFT and SEET derive from stationary embedding functionals evaluated with the self-consistent Green's function.
For an electronic Hamiltonian with an instantaneous bare Coulomb interaction, the total energy can be written schematically as
\begin{equation}
    E^{X}_{\mathrm{tot}}
    =
    E_{\mathrm{nn}}
    +
    \mathrm{Tr}[H_0G_{\mathrm{latt}}]
    +
    E_{\mathrm{H}}[\rho]
    +
    \tfrac{1}{2}\mathrm{Tr}[\Sigma_{\mathrm{xc}}G_{\mathrm{latt}}],
    \qquad
    \Sigma_{\mathrm{xc}}
    =
    \Sigma^{\mathrm{GW}}_{\mathrm{xc}}
    +
    \Sigma_{\mathrm{imp,xc}}
    -
    \Sigma_{\mathrm{DC,xc}}.
    \label{eq:etot-T1}
\end{equation}
Here, \(H_0\) contains the kinetic and external-potential terms, \(E_{\mathrm{nn}}\) is the nuclear repulsion, and \(E_{\mathrm{H}}\) is the Hartree energy of the interacting density \(\rho\).
The self-energy components exclude their Hartree contributions, and the impurity and double-counting terms are embedded into the lattice basis.
The diagrammatic subtraction removes the GW contribution replaced by the impurity treatment (Section~\ref{sec:beyond-dmft}).

These expressions give the internal energy.
At the operating temperature, thermodynamic stability instead depends on free-energy differences.
For DFT+DMFT, the electronic free energy can be evaluated with a stationary functional \citep{HauleBirol2015}.
The formal basis also exists for Tier~T1: GW+DMFT uses a functional of the Green's function and screened interaction \citep{Almbladh1999Psi,Biermann2003GWDMFT}, while SEET derives from a Luttinger-Ward functional of the Green's function \citep{ZgidGull2017}.
Their practical evaluation for the interface models considered here remains to be demonstrated.

Written with diagrammatic double counting, the GW+DMFT interaction functional has the schematic structure \(\Psi[G,W]=\Psi^{\mathrm{GW}}[G,W]+\Psi_{\mathrm{imp}}[G_{\mathrm{loc}},W_{\mathrm{loc}}]-\Psi^{\mathrm{GW}}_{\mathrm{loc}}[G_{\mathrm{loc}},W_{\mathrm{loc}}]\), with a sum over impurity regions where appropriate. SEET follows an analogous diagrammatic replacement construction, but uses a Luttinger--Ward functional \(\Phi[G]\) in which the weak-coupling contribution within each active space is replaced by the corresponding strongly correlated contribution \citep{Almbladh1999Psi,Biermann2003GWDMFT,ZgidGull2017}. Stationarity requires that the electronic self-energy and, for GW+DMFT, the polarization be obtained consistently from the chosen functional and that the corresponding self-consistency conditions be satisfied. Finite baths and truncated boson spaces introduce representation errors, but do not necessarily destroy functional consistency. Sampling noise and incomplete self-consistency introduce additional numerical errors. The effect of these approximations on stationarity and on the target free-energy differences must therefore be assessed. A prerequisite for the Tier~T1 thermodynamic calculations is to evaluate the embedding free energy consistently with the chosen numerical approximations.

The entropy has electronic, vibrational and configurational contributions, obtained from the embedding, classical phonon calculations and the defect-chemistry model of Section~\ref{sec:micro-inputs}, respectively.

The electronic contribution does not require a direct entropy measurement.
At the converged hybridization function, the impurity grand potential follows from the impurity partition function, normalized by that of the uncoupled bath.
Classical polynomial-time guarantees for this quantity assume a constant-size impurity coupled to a purely fermionic bath \citep{JiangEtAl2026,ArunachalamEtAl2026}; they do not establish efficiency as the active space grows.
For a static interaction, the impurity free-energy contribution can alternatively be obtained by coupling-constant integration of \(\langle H_{\mathrm{int}}\rangle\), starting from the known non-interacting reference while keeping the converged bath fixed.
This is one ingredient of the embedding free energy, whose remaining terms are evaluated classically.
Each integration point requires thermal-state preparation, so this route avoids real-time evolution but not the cost \(C_{\mathrm{Gibbs}}\) of Section~\ref{sec:qubits}.
The integration must be converged to the required accuracy.

Stationarity suppresses first-order errors from consistent variations of the functional's arguments around a self-consistent solution.
This property is established for the stationary DFT+DMFT functional \citep{HauleBirol2015} and requires a consistent, functional-derived formulation in Tier~T1.
It does not eliminate direct errors in the estimated impurity free energy or other functional terms.
Electronic entropy can be important at solid-oxide-cell temperatures, particularly when different atomic configurations favour different local spin states.
We retain \(E_{\mathrm{vac}}\) and \(E_{\mathrm{mig}}\) for energy differences and write \(\Delta G_{\mathrm{vac}}\) for the vacancy formation free energy, including the appropriate chemical reservoir (Section~\ref{sec:micro-inputs}).

We evaluate the energies separately for the four atomic configurations needed above: the pristine interface, the vacancy-containing interface, and the initial- and transition-state configurations of the migrating oxygen ion.
The resulting energies, \(E_{\mathrm{tot}}\), \(E_{\mathrm{tot}}^{V_{\mathrm O}}\), \(E_{\mathrm{IS}}\), and \(E_{\mathrm{TS}}\), enter Eqs.~\eqref{eq:evac} and \eqref{eq:emig}.

The interacting one-electron contribution can be written
\begin{equation}
    \mathrm{Tr}[H_{\mathrm{KS}}G_{\mathrm{latt}}]
    =
    -\frac{1}{\pi}
    \sum_{\mathbf{k}}w_{\mathbf{k}}
    \int d\omega\,f(\omega)\,
    \mathrm{Im}\,
    \mathrm{Tr}\!\left[
    H_{\mathrm{KS}}(\mathbf{k})
    G_{\mathrm{latt}}(\mathbf{k},\omega+i0^+)
    \right],
\end{equation}
where \(w_{\mathbf{k}}\) are the Brillouin-zone integration weights, \(f(\omega)\) is the Fermi--Dirac distribution, and \(\omega\) has units of energy with \(\hbar=1\).
In Tier~T1, the corresponding one-electron term uses \(H_0\).

For a static local interaction, the interaction energy can be obtained from the Galitskii--Migdal expression:
\begin{equation}
    \left\langle H_{\mathrm{int}}\right\rangle
    =
    -\frac{1}{2\pi}
    \sum_I
    \int d\omega\,f(\omega)\,
    \mathrm{Im}\,
    \mathrm{Tr}\!\left[
    \Sigma_{\mathrm{imp},I}(\omega+i0^+)
    G_{\mathrm{loc},I}(\omega+i0^+)
    \right].
    \label{eq:galitskii}
\end{equation}
Here, \(\Sigma_{\mathrm{imp},I}\) is the interaction self-energy, including its Hartree contribution, and \(G_{\mathrm{loc},I}\) is the lattice Green's function projected onto the correlated subspace at site \(I\).
At self-consistency, it equals the impurity Green's function.
For SEET, or GW+DMFT with a static reduction of the screened interaction, this expression gives the impurity interaction energy.
The impurity interaction energy can also be measured directly as the static thermal expectation value \(\mathrm{Tr}[\rho_\beta H_{\mathrm{int}}]\), avoiding the frequency integral and its high-frequency tail.
For dynamical screening represented by Eq.~\eqref{eq:hbos}, the corresponding energy estimator must be derived consistently from the electron--boson embedding formulation.

The quantum impurity solver provides \(G_{\mathrm{imp}}\), from which the impurity self-energy is extracted and embedded into the lattice calculation.
The lattice Green's function is then updated through the self-consistency loop until \(G_{\mathrm{imp}}\simeq G_{\mathrm{loc}}\).
GW+DMFT additionally requires bosonic self-consistency for the screened interaction.
A more detailed breakdown is provided in Section~\ref{sec:materials_to_impurity}.

\paragraph{Vacancy energetics as a function of \(\delta\).}
The thermodynamic factor used in Section~\ref{sec:micro-inputs} depends on how \(E_{\mathrm{vac}}\) changes with the vacancy concentration.
In LSCF the thermodynamic factor is large, and it is set by the \(\delta\) dependence of \(\mu_{\mathrm O}\), which is controlled by the hole states in the Co/Fe--O bands, that is, by correlation.
The embedding must therefore deliver \(E_{\mathrm{vac}}\) at several vacancy concentrations (supercells with one, two and three vacancies), and hence \(\mu_{\mathrm O}(\delta)\).
This requirement is also an opportunity: \(\delta(T,p_{\mathrm O_2})\), and therefore \(\mu_{\mathrm O}(\delta)\), is measured by thermogravimetry and coulometric titration \citep{Mizusaki1989LSC,Hashimoto2010LSCF}, which gives a direct experimental benchmark for the embedding at each active-space size.

\paragraph{Electronic conductivity.} We can derive the electronic conductivity in the interface region from the interacting spectral function:
\begin{equation}
    A (\mathbf{k},\omega) = -\frac{1}{\pi} \mathrm{Im} \left[ \omega+\mu-H_{\mathrm{KS}} (\mathbf{k})-\Sigma_\mathrm{latt} (\omega) \right]^{-1},
\label{eq:spectral}
\end{equation}
where \(\mu\) is the electronic chemical potential. In Tier T1, $H_{\mathrm{KS}}(\mathbf{k}) + \Sigma_{\mathrm{latt}}(\omega)$ is replaced by $H_0(\mathbf{k}) + \Sigma_{\mathrm{H}} + \Sigma^{\mathrm{GW}}_{\mathrm{xc}}(\mathbf{k},\omega) + \Sigma_{\mathrm{imp,xc}}(\omega) - \Sigma_{\mathrm{DC,xc}}(\omega)$, with the impurity matrices embedded into the lattice basis.

A Kubo-form conductivity estimate may then be written schematically as
\begin{equation}
    \sigma_e^{\mathcal I} \propto \int d\omega \left( -\frac{\partial f}{\partial \omega}\right) \sum_{\mathbf{k}} v_{\mathbf{k}}^2 A (\mathbf{k},\omega)^2.
\label{eq:kubo}
\end{equation}
Here, \(v_{\mathbf{k}}\) is the band velocity and \(-\partial f/\partial\omega\) restricts the dominant contribution to states near the Fermi level.
The spectral function enters squared because the Kubo-bubble conductivity is built from a current-current correlation function containing two Green's function propagators, each of which contributes a factor of \(A(\mathbf{k},\omega) \sim \mathrm{Im}\,G_\mathrm{latt}(\mathbf{k},\omega)\).
For several orbitals, \(v_{\mathbf k}^2 A^2\) stands for \(\mathrm{Tr}[v_{\mathbf k}A(\mathbf k,\omega)\,v_{\mathbf k}A(\mathbf k,\omega)]\), with \(v_{\mathbf k}\) the velocity matrices, and vertex corrections and polaronic contributions are neglected.
The conductivity checks whether the electrode satisfies the electronic-transport requirement of Section~\ref{sec:micro-inputs}.

\subsection{Additional material-screening criteria}
\label{other-material-screening-criteria}
This paper focuses on articulating a use case of quantum computing to improve materials screening based on more accurate computations of these interface observables, and in subsequent sections, we will see how these microscopic observables connect to an important macroscopic property of the electrode: the area-specific resistance (ASR).
However, it is worth noting that, while improving the accuracy of the ASR screen is very valuable, candidates passing the ASR screen must still be evaluated against a broader set of requirements, including thermal-expansion compatibility with the electrolyte and neighboring stack components; chemical and phase stability under the relevant temperature and oxygen chemical potential; interfacial reactivity and interdiffusion; mechanical integrity, fracture and delamination resistance; redox and thermal-cycling stability; cation segregation and long-term microstructural evolution; chromium and other contaminant tolerance; catalytic stability under both SOFC and SOEC operation; gas-transport limitations in the realized porous electrode; electronic and ionic percolation after processing and ageing; sinterability and compatibility with the fabrication route; and material abundance, cost, toxicity, and supply-chain constraints \cite{NisarEtAl2024}.
Thus, the quantum-enabled ASR model presented in this paper should be interpreted as a high-value screening layer within a multi-criteria materials qualification process rather than as a complete electrode digital twin.

We note that phase stability is itself a potential extension of the quantum-enabled GW+DMFT workflow when the competing phases contain strongly correlated electronic states.
One can compute correlated free energies for the candidate material and for competing crystal structures, and possible reaction or decomposition products using GW+DMFT. Comparing these energies as a function of composition, temperature, oxygen chemical potential, and strain then indicates which phase is thermodynamically favored under given operating conditions \cite{ParkEtAl2014,IsaacsMarianetti2020}.

\subsection[Choice of system and motivation for correlated embedding]{Choice of system and motivation for correlated embedding}
\label{sec:motivation}
We use the LSCF/GDC interface as the running example because the literature covers both the electrode composition and the interface.
The composition \(\mathrm{La_{0.6}Sr_{0.4}Co_{0.2}Fe_{0.8}O_{3-\delta}}\) is the one studied by \citet{AsakuraEtAl2026LSCFGDC}.
LSCF/GDC interfaces have been observed to show interdiffusion, oxygen-vacancy ordering, interfacial strain, and changes in transition-metal valence and spin state \citep{LiEtAl2011LSCFGDC,LeeEtAl2016LSCFHeterostructure,AsakuraEtAl2026LSCFGDC}, and \citet{LeeEtAl2016LSCFHeterostructure} report the orientation relationship adopted in Section~\ref{sec:materials_to_impurity}.
The measured bulk oxygen non-stoichiometry of LSCF \citep{Hashimoto2010LSCF} provides a first experimental benchmark (Section~\ref{sec:observables}).

Vacancy energetics in these perovskites are usually computed with DFT or DFT+\(U\), and the results depend strongly on the structural model, the magnetic order, and the Hubbard correction.
For \(\mathrm{La_{1-x}Sr_xCo_{0.25}Fe_{0.75}O_{3-\delta}}\), \citet{Gryaznov2012LSCF} find spreads of up to about 1.7~eV relative to thermogravimetric data, and for transition-metal-doped \(\mathrm{La_{0.5}Sr_{0.5}CoO_3}\), \citet{Wei2025LSCF} find differences of 0.3--0.6~eV between ferromagnetic and paramagnetic states.
These studies do not isolate the contribution of electronic correlation, and we do not read them as evidence for it; they show only that DFT-level vacancy energies are sensitive to modeling choices.
We therefore treat correlated embedding as a motivated hypothesis for this system, not as a demonstrated necessity.
Correlation can enter the electrode metrics through three channels, each already present in the workflow.
 
\paragraph{Thermodynamic factor.}
The thermodynamic factor \(\gamma_{\mathrm O}\) is set by the \(\delta\) dependence of \(\mu_{\mathrm O}\), which the hole states in the Co/Fe--O bands control (Section~\ref{sec:observables}).
Together with \(E_{\mathrm{vac}}\), \(\gamma_{\mathrm O}\) determines the chemical diffusion coefficient and therefore the ASR (Section~\ref{sec:micro-inputs}).
\citet{Augustinsky2013LaCoO3} report that correlation effects in \(\mathrm{LaCoO_3}\) grow with Sr doping, that is, with the hole doping that sets \(\gamma_{\mathrm O}\).
 
\paragraph{Double counting.}
Standard double-counting prescriptions shift the relative position of the transition-metal \(d\) and O \(2p\) states, and with it the energetics \citep{Dang2014Covalency,Haule2015}.
This spread is intrinsic to the method and may be large enough to change a candidate ranking (Section~\ref{sec:amplification}).
The evidence of \citet{Dang2014Covalency} concerns titanates and vanadates, so its transfer to cobaltite--ferrites is an analogy, not a result. Tier T1 removes the free parameter of the double counting (Section~\ref{sec:beyond-dmft}), which is the main motivation for using it; residual dependence on projectors, active space and the GW starting point remains. Whether the residual spread is smaller than the double-counting spread is a hypothesis. It can be tested on bulk LSCF, alongside the bulk benchmark of Section~\ref{sec:tiers}, by comparing the spread of the Tier T1 results over projector and GW-starting-point choices with the spread of DFT+DMFT over standard double-counting prescriptions.
 
\paragraph{Spin-state entropy.}
At operating temperature, changes in the Co spin state contribute an entropy term that zero-temperature DFT+\(U\) does not capture.
Our estimate of this term is 60--90~meV, and \citet{Chakrabarti2016LaCoO3} report about 110~meV per formula unit in \(\mathrm{LaCoO_3}\).
Because the term enters the exponent of the vacancy concentration, it changes \(c_{\mathrm{vac}}\) by a factor of \(\exp(\Delta E/k_BT)\approx2\)--\(3\) for \(\Delta E=60\)--\(90\)~meV at 973~K.
 
\paragraph{Limitations.}
Three limitations qualify this motivation.
First, Co occupies only 20\% of the \(B\) sites in our example, whereas most of the evidence above concerns Co-rich systems.
Second, DMFT does not address structural and sampling errors, which reach about 1~eV from the choice of supercell alone \citep{Gryaznov2012LSCF}.
Third, whether classical impurity solvers actually limit the calculation at the required active-space size remains open; the comparison with the best classical impurity solvers (Section~\ref{sec:tiers}) is designed to test it.

\section{From candidate materials to a quantum impurity Hamiltonian}
\label{sec:materials_to_impurity}

This section defines the classical translation procedure that converts a candidate electrode/electrolyte interface into the Anderson impurity Hamiltonians, one per correlated site and per atomic configuration, required by a quantum impurity solver. The sequence is
\[
\boxed{
\begin{array}{c}
\text{electrode and electrolyte crystals}
\\
\downarrow
\\
\text{atomistic interface geometry}
\\
\downarrow
\\
\text{periodic DFT calculation}
\\
\downarrow
\\
\text{(Tier T1) GW calculation: } \Sigma^{\mathrm{GW}}(\mathbf{k},\omega),\ W(\omega)
\\
\downarrow
\\
\text{localized lattice Hamiltonian}
\\
\downarrow
\\
\text{correlated local subspace}
\\
\downarrow
\\
\text{DMFT hybridization function}
\\
\downarrow
\\
\text{finite Anderson impurity Hamiltonian}.
\end{array}
} 
\]

DFT treats the full periodic interface
supercell. 
 The embedding subsequently selects local correlated orbitals from this DFT-derived lattice model. 
Each impurity solver remains local: the effect of the surrounding electrode, electrolyte, vacancies, and interface is encoded in that site's local one-electron Hamiltonian and hybridization function. Because the interface breaks the symmetry that would otherwise make correlated sites equivalent, multiple, symmetry-inequivalent sites each require their own impurity solver.

\subsection{Construction of the atomistic interface}
For each electrode/electrolyte pair, experimentally known parent crystal structures are expanded into compositionally appropriate supercells, dopants and oxygen vacancies are assigned to explicit sites, and candidate interfaces are generated by enumerating crystallographic orientation, termination, lateral registry, and low-strain commensurate surface cells.
Each candidate is subsequently relaxed with periodic DFT and ranked by interface energy, with lower values indicating more energetically favorable interfaces. Full crystallographic coordinates and construction details are provided in Appendix~\ref{app:crystallography}.

For the running LSCF/GDC example, we use 
\[
\mathrm{La_{0.6}Sr_{0.4}Co_{0.2}Fe_{0.8}O_{3-\delta}} \;|\; \mathrm{Ce_{0.9}Gd_{0.1}O_{1.95}}
\]
(i.e.~\(x=0.4\), \(y=0.8\) in our general LSCF formula above), where \(\delta\) denotes the oxygen non-stoichiometry, with the experimentally motivated
\[
(001)_{\mathrm{LSCF,pc}}\parallel(001)_{\mathrm{GDC,c}},
\qquad
[100]_{\mathrm{LSCF,pc}}\parallel[110]_{\mathrm{GDC,c}}
\]
orientation \cite{LeeEtAl2016LSCFHeterostructure}, where \((hkl)\) and \([uvw]\) denote crystallographic planes and directions, respectively, specifying which faces of the two materials are joined and their relative in-plane rotational alignment, and \(\mathrm{pc}\) and \(\mathrm{c}\) denote the pseudocubic perovskite and cubic fluorite settings of LSCF and GDC, respectively.
Candidate structures spanning different terminations, cation arrangements, oxygen-vacancy positions, and lateral registries are then relaxed and compared by interface energy, as described above, before the correlated electronic-structure calculation.

\subsection{From the atomistic interface to the quantum impurity problem}
\label{sec:interface_to_impurity_summary}
The purpose of the classical embedding preprocessing is to translate a candidate electrode/electrolyte interface into a set of local many-body problems that can be solved on a quantum computer. 
The detailed crystallographic construction and the vacancy and migration geometries are given in Appendices \ref{app:crystallography} and \ref{app:select-vacancy}.
The remaining DMFT equations are standard \citep{GeorgesEtAl1996,KotliarEtAl2006} and are implemented in codes such as TRIQS/DFTTools \citep{AichhornEtAl2016}.
In the GW-based embedding of Tier T1 (Section~\ref{sec:tiers}), a GW calculation supplies the non-local self-energy and the double counting is the local part of the GW self-energy (Section~\ref{sec:beyond-dmft}).
GW+DMFT also subtracts the local polarization and passes $U(\omega)$ to the impurity, while SEET uses the bare interaction within the active space.
Here we articulate the resulting computational objects and the division of labor between the classical and quantum parts of the workflow.
 
For each candidate electrode/electrolyte pair, a relaxed, screened atomistic interface supercell provides the starting point for the correlated electronic-structure calculation.
Oxygen vacancy formation requires calculations for both the pristine and vacancy-containing structures, while vacancy migration additionally requires initial- and transition-state geometries, with the transition state obtained from a nudged-elastic-band (NEB) calculation \citep{HenkelmanEtAl2000}.

Periodic DFT provides the Kohn-Sham Hamiltonian $H_\mathrm{KS}$ for every retained atomic geometry; its matrix elements in a localized Wannier/projected-orbital basis define the one-electron lattice Hamiltonian $H^0(\mathbf{k})$ \citep{MarzariEtAl2012,AichhornEtAl2016}, from which a subset of strongly correlated, symmetry-inequivalent sites $I$ is selected for explicit many-body treatment (for LSCF/GDC: the Fe and Co $3d$ shells, and Ce $4f$ where relevant).
Because the choice of localized orbitals can change DFT+DMFT results quantitatively \citep{KarpEtAl2021}, the localized model is accepted only if it reproduces the DFT bands and orbital characters over the energy window used.
Each site contributes a local one-electron Hamiltonian $h_I$ and screened interaction parameters $U_I, J_I$ that parameterize $H_{\mathrm{int}}^I$, computed for example with the constrained random-phase approximation \citep{AryasetiawanEtAl2004} (reference DFT+DMFT form).
In Tier T1, the interaction is $U_I(\omega)$ derived from the GW screened interaction in GW+DMFT, or the bare interaction within the active space in SEET.
Summed over all sites, these define the formal interacting lattice Hamiltonian underlying DMFT, analogous to Eq.~\eqref{eq:etot} but resolved per site, not solved directly as a full many-body matrix.
 
DMFT instead maps each correlated site $I$ onto an Anderson impurity model \citep{GeorgesEtAl1996,KotliarEtAl2006}.
The remainder of the lattice enters that local problem through the frequency-dependent hybridization function $\Delta_I(z)$, where $z$ is a general complex frequency, discretized into a finite bath (energies $\{\epsilon_\ell^I\}$, couplings $\{V_{m\ell}^I\}$) that, together with $h_I$, $U_I$, and $J_I$ (for GW+DMFT, the bare interaction plus the bosons of Eq.~\eqref{eq:hbos}), defines the finite Anderson impurity Hamiltonian $H_{\mathrm{AIM},I}$,
\begin{equation}
    H_{\mathrm{AIM},I}
    = \underbrace{\sum_{mm'\sigma} (h_I)_{mm'}\, d^{\dagger}_{Im\sigma} d_{Im'\sigma}
    + \sum_{\ell\sigma} \epsilon_{\ell}^{I}\, b^{\dagger}_{I\ell\sigma} b_{I\ell\sigma}
    + \sum_{m\ell\sigma} \left( V_{m\ell}^{I}\, d^{\dagger}_{Im\sigma} b_{I\ell\sigma} + \mathrm{h.c.} \right)}_{H_{\mathrm{quad},I}}
    + H_{\mathrm{int}}^{I},
\label{eq:aim}
\end{equation}
where \(d_{Im\sigma}\) and \(b_{I\ell\sigma}\) annihilate an electron with spin \(\sigma\) in impurity orbital \(m\) and bath orbital \(\ell\), respectively.
For GW+DMFT with dynamical screening, $H_{\mathrm{AIM},I}$ acquires auxiliary bosonic modes,
\begin{equation}
H_{\mathrm{bos},I} = \sum_\nu \omega_\nu a^\dagger_\nu a_\nu
 + \sum_\nu g_\nu \left(a_\nu + a^\dagger_\nu\right)
 \left(\hat N_I - N_0\right),
\label{eq:hbos}
\end{equation}
whose parameters $\{\omega_\nu, g_\nu\}$ reproduce the frequency dependence of $U_I(\omega)$ (density-density screening).
This captures the uniform (monopole) part of the screening only.
$J$ is taken as static, from the cRPA value $J(0)$, and both $J(\omega)$ and the orbital dependence of $U(\omega)$ are neglected.
A classical check bounds the effect of this choice: compute the cRPA $J(\omega)$ for LSCF, and the sensitivity of the Co spin-state energy difference to $J\pm20\%$.
In this case, $H^I_{\mathrm{int}}$ carries the high-frequency limit $U_I(\omega\to\infty)$, the bare Coulomb interaction of the correlated orbitals, and the bosons screen it to $U_I(0) = U_I(\infty) - \sum_\nu 2g_\nu^2/\omega_\nu$ at low frequency \citep{WernerMillis2010DynamicalScreening}.
Putting the screened interaction in $H^I_{\mathrm{int}}$ would screen twice.
We couple the bosons to $\hat N_I - N_0$, with $N_0$ the mean impurity filling, so that the modes are displaced only by charge fluctuations and the boson truncation stays small.
The difference is a one-body term absorbed into $\mu$.

SEET uses Eq.~\eqref{eq:aim} with the bare interaction and no bosons.
GW+DMFT with a static reduction of $U(\omega)$ uses it with the screened static value and a renormalized bandwidth.

The number of bath orbitals \(N_{\mathrm{bath}}\) is increased until the target observables converge.
The discretization is a nonlinear fit of \(\Delta_I\) on the Matsubara axis at the operating temperature.
The fit has many near-degenerate local minima and the fitted bath changes between iterations, so the observables need not converge monotonically with \(N_{\mathrm{bath}}\).
The convergence of \(E_{\mathrm{vac}}\) and \(E_{\mathrm{mig}}\) with \(N_{\mathrm{bath}}\) must therefore be reported explicitly, since bath discretization error enters the total energies directly and does not cancel between configurations that have different hybridization functions.
 
This is the point at which the quantum computer enters the workflow.
Rather than simulating the complete electrode/electrolyte supercell, the quantum processor solves each local interacting Hamiltonian $H_{\mathrm{AIM},I}$ and returns the corresponding finite-temperature impurity Green's function $G_{\mathrm{imp},I}(z)$ (Section~\ref{sec:quantum-solver}). Via Dyson's equation, $G_{\mathrm{imp},I}(z)$ yields the local self-energy $\Sigma_I(z)$, the central many-body correction, local but frequency-dependent, which is embedded back into the lattice and iterated to DMFT self-consistency, $G_{\mathrm{imp},I}(z) \simeq G_{\mathrm{loc},I}(z)$ (in fully self-consistent GW+DMFT, also $W_{\mathrm{imp},I} \simeq W_{\mathrm{loc},I}$).

At convergence, the calculation therefore provides, for every inequivalent correlated site,
\[
\left\{
G_{\mathrm{imp},I}(z),
\Sigma_I(z),
\Delta_I(z)
\right\},
\]
which are the local correlated quantities required to reconstruct the interacting electronic structure of the full interface. 
 
These local self-energies are embedded into the full Wannier lattice basis to reconstruct the interacting lattice self-energy $\Sigma_\mathrm{latt}(\mathbf{k},z)$ and Green's function $G_\mathrm{latt}(\mathbf{k},z)$ \citep{KotliarEtAl2006}, which are then used to compute $E_\mathrm{vac}^{\mathcal I}$, $E_\mathrm{mig}^{\mathcal I}$, and $\sigma_e^{\mathcal I}$ as described in Section~\ref{sec:observables} above.
 
\subsubsection[The quantum impurity solver]{The quantum impurity solver}
\label{sec:quantum-solver}
At each DMFT iteration, the quantum computer receives \(H_{\mathrm{AIM},I}\) and returns its finite-temperature Green's function.
The impurity and bath spin orbitals are first mapped to qubits (Section~\ref{sec:qubits}).
The solve then has three steps: thermal-state preparation, real-time evolution, and measurement of the particle and hole correlation functions.
 
\paragraph{Thermal-state preparation.}
Because the cell operates at finite temperature, the target is the thermal state of the impurity model rather than only its ground state:
\begin{equation}
    \rho_{\beta} = \frac{e^{-\beta(H_{\mathrm{AIM},I}-\mu \hat N)}}{Z},
    \qquad
    Z = \mathrm{Tr}\, e^{-\beta(H_{\mathrm{AIM},I}-\mu \hat N)},
    \qquad
    \beta = \frac{1}{k_B T},
\label{eq:rho-beta}
\end{equation}
where \(\hat N\) is the electron-number operator and \(\mu\) the electronic chemical potential.
The target is \(\rho_\beta\) of Eq.~\eqref{eq:rho-beta} at \(\beta\approx 12\)~eV\(^{-1}\).
Fault-tolerant routes include Gibbs samplers built from quantum Markov chains that satisfy detailed balance, whose cost is set by the mixing time~\cite{ChenEtAl2023GibbsSampler}.
The quadratic part \(H_{\mathrm{quad},I}\) is diagonalized classically, so its Gibbs state, a fermionic Gaussian state, can be prepared exactly and used to start the chain.
If bosonic modes are present, their free thermal state is included in this reference.
\(C_{\mathrm{Gibbs}}\) enters every sample (Section~\ref{sec:qubits}).
For a constant-size impurity coupled to a purely fermionic bath, Jiang et al.\ give a quantum algorithm that prepares the thermal state \citep{JiangEtAl2026}, in contrast to most Gibbs samplers, whose runtime bounds are exponential or worse in the inverse temperature.
The runtime is polynomial in the system size, the inverse temperature, the bath energy scale and the inverse precision only for a constant impurity size: the runtime depends exponentially on the number of impurity modes, so the guarantee does not establish efficiency for enlarged active spaces, and it does not cover bosonic modes.
For the enlarged active spaces and the bosonic modes of Tier T1, the thermal-state cost, through the mixing time or an equivalent bound, remains open (Section~\ref{sec:open}).
Quantum imaginary-time evolution~\cite{MottaEtAl2020} is a near-term heuristic and gives no fault-tolerant cost guarantee.
 
\paragraph{Real-time evolution and measurement.}
From the thermal state, the quantum computer applies the controlled Majorana operators of an interferometric circuit, evolves under \(H_{\mathrm{AIM},I}\), and measures the ancilla to obtain the correlation functions
\begin{equation}
    G^{>}_{mm'}(t) = -i\,\mathrm{Tr}\!\left[\rho_\beta\, d_m(t)\, d_{m'}^\dagger\right],
    \qquad
    G^{<}_{mm'}(t) = i\,\mathrm{Tr}\!\left[\rho_\beta\, d_{m'}^\dagger\, d_m(t)\right],
    \qquad
    d_m(t) = e^{iH_{\mathrm{AIM},I}t}\, d_m\, e^{-iH_{\mathrm{AIM},I}t}.
\end{equation}
These correlators give the retarded Green's function, \(G^{R}(t)=\theta(t)\left[G^{>}(t)-G^{<}(t)\right]\), and by Fourier transformation the spectral function \(A_I(\omega)\) on the real-frequency axis.
Because \(d_m\) is not unitary, the correlators are measured through the unitary Majorana combinations \(d_m+d_m^\dagger\) and \(i(d_m-d_m^\dagger)\) with an interferometric (Hadamard-test) circuit, as in an early quantum-DMFT proposal \citep{BauerEtAl2016}.
For GW+DMFT with the monopole screening of Eq.~\eqref{eq:hbos}, the solver must also return the charge correlator \(\langle \hat N_I(t)\hat N_I(0)\rangle\), needed for the bosonic self-consistency \(W_{\mathrm{imp},I}\simeq W_{\mathrm{loc},I}\); since \(\hat N_I\) is a sum of mode occupations, i.e.~ of Pauli-\(Z\) operators under the Jordan--Wigner mapping, it is measured term by term with the same type of circuit.
Orbital-resolved screening would require orbital-resolved density correlators.
The Matsubara Green's function used in the self-consistency loop follows from the spectral representation, \(G_{\mathrm{imp},I}(i\omega_n)=\int d\omega\, A_I(\omega)/(i\omega_n-\omega)\), without analytic continuation.
 
\paragraph{Sampling noise and self-consistency.}
Each correlator is estimated from a finite number of shots at a finite set of times, with a maximum evolution time \(t_{\max}\).
The Fourier transform to the real-frequency axis is therefore truncated, and its resolution and ringing must be controlled by windowing and checked against sum rules.
Dyson's equation, \(\Sigma=\mathcal{G}_0^{-1}-G_{\mathrm{imp}}^{-1}\), inverts \(G_{\mathrm{imp}}\) and so amplifies noise and truncation artefacts where the spectral weight is small, and the subsequent bath refit adds its own ill-conditioning.
The self-consistent iterates then fluctuate around the fixed point instead of converging to it, as in CT-QMC-based DMFT, and the converged quantities must be averaged over the last iterations.
The relevant precision is that of the energy differences \(E_{\mathrm{vac}}\) and \(E_{\mathrm{mig}}\), not of \(G\) itself.
Bauer et al.\ chose the number of measurements so that the sampling error is small compared with the uncertainty of the bath fit \citep{BauerEtAl2016}, which is a sensible rule but not a bound for energy differences.
The propagation of correlator noise to these energy differences is the precision budget of Section~\ref{sec:open}.

\paragraph{Classical alternatives.}
Two concurrent preprints, Jiang et al.~\cite{JiangEtAl2026} and Arunachalam et al.~\cite{ArunachalamEtAl2026}, give polynomial-time classical algorithms for static properties of quantum impurity models with a constant-size impurity and a free-fermion bath: the partition function, and hence the free energy, and thermal expectation values of static observables (fermionic Gaussian observables in Jiang et al., sums of polynomially many Majorana monomials in Arunachalam et al.), at any temperature.
The DMFT loop needs more: the impurity Green's function is a two-time correlator, on the imaginary or the real time axis, and is not covered by these results.
Jiang et al.\ conclude that a superpolynomial quantum advantage for quantum-enhanced DMFT, if any, must come from such dynamical quantities.
Their hardness result concerns nonequilibrium Green's functions of time-dependent impurity Hamiltonians, not the equilibrium Green's function of a static \(H_{\mathrm{AIM},I}\), whose classical hardness is open.
Arunachalam et al.\ show that estimating a mode occupation after real-time evolution under a time-independent impurity Hamiltonian with a constant-size impurity is BQP-complete, but for a specially prepared Slater-determinant initial state and a bath constructed for the purpose, not for a thermal state of a fitted DMFT bath.
Neither result therefore decides the hardness of the equilibrium Green's function at the operating temperature, which is why the comparison with the best classical impurity solvers is an empirical test and not a theorem.
The classical runtimes of both works are also exponential in the number of impurity orbitals and assume a fermionic bath, which is why the impurity-solver requirements discussed in Section~\ref{sec:quantum-algos} matter.

\subsection{Logical-qubit requirements and computational cost drivers}
\label{sec:qubits}

The relevant resource target is an impurity model whose predictions of vacancy and migration energetics are converged. Its quantum cost depends on the interacting active space, the bath representation, and the precision required to distinguish candidate materials.

For \(N_{\mathrm{imp}}\) spatial impurity orbitals and \(n_b\) spatial bath orbitals per impurity orbital, a Jordan--Wigner or Bravyi--Kitaev encoding requires
\begin{equation}
    N_q^{(\mathrm{ferm})}
    =
    2(N_{\mathrm{imp}}+N_{\mathrm{bath}})
    =
    2N_{\mathrm{imp}}(1+n_b)
    \label{eq:qubits}
\end{equation}
logical register qubits, before symmetry reductions and ancillas. Table~\ref{tab:qubits} illustrates the register sizes for a single transition-metal \(d\) shell and a two-site cluster. These counts describe encoding requirements; the bath size must be established by convergence of the target observables.

\begin{table}[t]
\centering
\caption{Illustrative fermionic register sizes versus the number \(n_b\) of spatial bath orbitals per impurity orbital. Ancillas are excluded.}
\label{tab:qubits}
\begin{tabular}{lrrrrr}
\toprule
\(n_b\) & 2 & 5 & 10 & 20 & 50 \\
\midrule
Single \(d\) shell, \(N_{\mathrm{imp}}=5\) & 30 & 60 & 110 & 210 & 510 \\
Two-site cluster, \(N_{\mathrm{imp}}=10\) & 60 & 120 & 220 & 420 & 1020 \\
\bottomrule
\end{tabular}
\end{table}

For dynamical screening, \(N_{\mathrm{bos}}\) binary-encoded bosonic modes truncated at \(n_{\max}\) quanta add \(N_{\mathrm{bos}}\lceil\log_2(n_{\max}+1)\rceil\) register qubits. The screening representation and occupation cutoff must likewise be converged.

For efficiently represented discretized impurity Hamiltonians, quantum time evolution can be implemented with gate cost polynomial in system size and evolution time, at fixed simulation accuracy and bounded Hamiltonian parameters. This provides a route to treating interacting active spaces beyond the practical reach of classical solvers. Whether it yields an overall advantage depends on thermal-state preparation, measurement precision, and repeated embedding solves.

For the direct-sampling protocol of Section~\ref{sec:quantum-solver}, the logical gate cost of one impurity solve is schematically
\begin{equation}
    C_{\mathrm{solve}}
    \approx
    N_{\mathrm{corr}}N_{\mathrm{shot}}
    \sum_{j=1}^{N_t}
    \left[C_{\mathrm{Gibbs}}+C_{\mathrm{evol}}(t_j)\right],
\end{equation}
where \(N_{\mathrm{corr}}\) counts the measured correlators, \(N_t\) the sampled times, and \(N_{\mathrm{shot}}\) the repetitions per correlator and time. Thermal preparation is repeated for each sample. Direct sampling has inverse-square dependence on additive correlator precision, while the evolution cost depends on the simulation method; block-encoding approaches depend on both the Hamiltonian normalization and the gate cost per query. The required precision and time range must be set by their propagated effect on vacancy and migration energies.

The full calculation multiplies this cost by the number of geometries, quantum-treated impurity regions, and embedding iterations. Classical solvers can be retained for tractable regions, concentrating quantum resources on the impurity problems that limit predictive accuracy. The classical GW and embedding workload must also be included when assessing the overall benefit.

These considerations define a concrete route to a resource estimate: establish the smallest converged impurity representation, determine the measurement accuracy needed for reliable material ranking, and evaluate the resulting preparation and simulation circuits. Comparing that workload with the best applicable classical workflow would establish whether the quantum extension is advantageous for this application. Physical-qubit requirements and runtime additionally require a specified fault-tolerant architecture (Section~\ref{sec:open}).

\subsection{Operational summary}
\label{sec:tiers}
The calculations are organized as a tiered screening funnel (Table~\ref{tab:tiers}): DFT screens all candidates, and the GW-based embedding with the quantum impurity solver is applied to the shortlist.

\begin{table}[t]
\centering
\caption{Tiered screening funnel.}
\label{tab:tiers}
\begin{tabular}{l p{0.36\textwidth} p{0.48\textwidth}}
\toprule
Tier & Method & Applied to \\
\midrule
T0 & DFT & all candidates: geometries, interfaces, NEB paths \\
T1 & GW-based embedding (GW+DMFT or SEET), quantum impurity solver; option selected by the bulk pilot (the embedding-method comparison on bulk LSCF, below) & shortlist: active spaces beyond classical reach; convergence of energy differences with active-space size \\
\bottomrule
\end{tabular}
\end{table}

Steps 1 to 5 below form Tier T0 and steps 6 to 16 form Tier T1. Steps 7 to 16 are written in the familiar DFT+DMFT form; in the GW-based embedding used here, a GW calculation on the T0 geometry supplies the non-local self-energy (step 6), and step 9 uses the diagrammatic DC instead of a DFT+DMFT prescription and the interaction is $U(\omega)$ (GW+DMFT) or the bare active-space interaction (SEET); step 14 uses the quantum impurity solver.
For each pristine, vacancy, or migration geometry, the complete translation procedure is:
\begin{enumerate}
    \item \textbf{(T0)} obtain the experimentally relevant bulk crystal structures;
    \item enumerate orientations, terminations, commensurate cells, and interface registries;
    \item assign dopants and alloy occupations explicitly;
    \item relax and rank the candidate interfaces with periodic DFT;
    \item enumerate chemically distinct oxygen-vacancy sites and migration paths;
    \item \textbf{(T1)}compute the GW self-energy $\Sigma^{\mathrm{GW}}(\mathbf{k},\omega)$ and the screened interaction $W(\omega)$ on the T0 geometry;
    \item \textbf{(T1)} construct a localized DFT lattice Hamiltonian containing the transition-metal \(d\), oxygen \(p\), and any relevant electrolyte states;
    \item select full local Fe and Co \(3d\) shells as the initial correlated impurity subspaces;
    \item compute the impurity interaction ($U(\omega)$ or the bare active-space interaction) and the diagrammatic double-counting term;
    \item initialize the lattice self-energy and project the lattice Green's function onto each impurity;
    \item obtain the continuous hybridization function;
    \item fit that function to finite bath energies and couplings;
    \item generate one Anderson impurity Hamiltonian for every inequivalent correlated site;
    \item encode and solve these Hamiltonians on the quantum computer to obtain the impurity Green's functions and self-energies;
    \item embed the self-energies into the full lattice basis and iterate the embedding loop to self-consistency, including the GW update if fully self-consistent;
    \item reconstruct the interacting lattice Green's function used to evaluate the macroscopic observables.
\end{enumerate}

The resulting data flow is:
\[
\boxed{
\begin{array}{c}
\mathrm{LSCF|GDC\ atomic\ geometry}
\\
\downarrow
\\
H^0(\mathbf{k}),\ \Sigma^{\mathrm{GW}}(\mathbf{k},\omega),\ U(\omega)
\\
\downarrow
\\
\left\{
h_I,\,
U_I,\,
J_I,\,
\Delta_I(i\omega_n)
\right\}
\\
\downarrow
\\
\left\{
\epsilon_{\ell}^{I},\,
V_{m\ell}^{I}
\right\}
\\
\downarrow
\\
H_{\mathrm{AIM},I}
\\
\downarrow
\\
G_{\mathrm{imp},I}
\\
\downarrow
\\
\Sigma_{\mathrm{latt}}(\mathbf{k},z),\ G_{\mathrm{latt}}(\mathbf{k},z)
\end{array}
}
\]

Tier~T1 targets enlarged correlated active spaces, potentially including transition-metal \(d\) shells and neighbouring O \(2p\) orbitals or clusters around a vacancy. GW+DMFT and SEET should first be benchmarked on bulk LSCF against available thermodynamic data and classical impurity calculations. The potential quantum advantage arises if convergence of \(E_{\mathrm{vac}}\) and \(E_{\mathrm{mig}}\) requires active spaces beyond the practical capabilities of the best classical solvers. The embedding method and quantum solver must therefore be assessed jointly in terms of predictive accuracy and total computational cost.

\paragraph{Scope of the GW step.}
The GW workload depends on the chosen level of self-consistency and the interface size.
Its computational cost and its contribution to the total workload must be benchmarked for the selected implementation.

\subsection{Dependency on the atomistic geometry}
\label{sec:geometry}
The accuracy of the embedding calculation is conditional on the quality of the atomistic geometry supplied to it. 
In the baseline workflow, interface structures, vacancy configurations, and migration pathways are first relaxed at the DFT level and are then held fixed while the correlated electronic problem is solved. 
If DFT predicts an incorrect local minimum, bonding geometry, vacancy configuration, or migration transition state, improving the many-body treatment at that fixed geometry does not remove the underlying structural error: the calculation would simply provide a more accurate electronic solution for the wrong atomic configuration. 
The atomistic relaxation should therefore be regarded as a separate source of model uncertainty from the many-body solution itself.
In Tier T1, DFT geometries combined with beyond-DFT energies remain the main residual inconsistency, most relevant where changes of the Co spin state alter ionic radii and hence bond lengths.

When structural forces are themselves sensitive to electronic correlations, the relaxation can be promoted from DFT to charge-self-consistent DFT+DMFT.
Analytical force formalisms have been developed in which the DFT+DMFT free energy is differentiated with respect to the atomic coordinates, allowing correlated materials to be structurally optimized rather than evaluating DMFT only on a fixed DFT geometry \cite{HauleBirol2015,HaulePascut2016,PlekhanovEtAl2021}. These force formalisms exist for DFT+DMFT. To our knowledge, no equivalent is yet available for GW+DMFT or SEET at this system size, so Tier T1 uses DFT geometries.
In such a calculation, the electronic and structural problems form an additional outer loop:
\[
\mathbf R^{(n)} \longrightarrow \mathrm{DFT+DMFT} \longrightarrow \mathbf F^{(n)} = -\frac{\partial F_{\mathrm{DFT+DMFT}}} {\partial \mathbf R} \longrightarrow \mathbf R^{(n+1)},
\]
where \(\mathbf{R}^{(n)}\) denotes the atomic coordinates at relaxation iteration \(n\), \(F_{\mathrm{DFT+DMFT}}\) is the correlated DFT+DMFT free energy, and \(\mathbf{F}^{(n)}\) denotes the corresponding atomic forces. The outer loop is repeated until both the electronic DMFT solution and the atomic forces are converged.

Therefore, a second, distinct quantum-computing opportunity arises if the atomic relaxation itself requires a correlated electronic treatment. 
DFT+DMFT structural optimization requires repeated evaluation of correlated free energies or forces as the nuclear coordinates are updated. 
In principle, a quantum impurity solver could replace the classical impurity solver within this outer DFT+DMFT force loop, so that the same quantum many-body subroutine used for the observables also contributes to the forces determining the geometry.

More generally, quantum algorithms for electronic-structure energy derivatives and molecular forces have already been proposed, including phase-estimation- and Hamiltonian-simulation-based approaches suitable for geometry optimization and molecular dynamics \cite{OBrienEtAl2019Derivatives,OBrienEtAl2022Forces}. 
This represents a separate potential quantum advantage from the impurity-solver acceleration studied in the baseline workflow. We do not include this additional structural relaxation cost or benefit in the application presented here.

\section{From microscopic observables to engineering metrics}
\subsection{Finite-thickness MIEC oxygen-electrode model}\label{sec:meic-model}
The electrode micromodel is selected according to the charge-transport and reaction architecture of the electrode rather than being universal across solid-oxide-cell chemistries. 
Two-phase electrodes such as Ni--YSZ and conventional LSM--YSZ are naturally described by composite-electrode frameworks such as Costamagna, whereas single-phase MIEC oxygen electrodes such as LSCF are more naturally represented by ALS-type models; composite MIEC/ionic-conductor electrodes require corresponding extensions \cite{Costamagna1998MicroModelling,AdlerLaneSteele1996,MortensenEtAl2014}.

Because the LSCF oxygen electrode is a mixed ionic--electronic conductor, we describe its polarization using the Adler--Lane--Steele (ALS) continuum framework \cite{AdlerLaneSteele1996,PanissetEtAl2023}. 
In this description, oxygen is incorporated by surface exchange at the internal gas/MIEC surface and subsequently transported through the solid phase. 
Electrode polarization therefore emerges from the coupled effects of oxygen surface exchange and solid-state oxygen transport.

For a macro-homogeneous porous MIEC electrode, neglecting gas-phase transport
limitations, the characteristic chemical resistance and time constant may be
written as \cite{Adler1998,PanissetEtAl2023}
\begin{equation}
    R_{\mathrm{chem}} = \frac{R_gT}{4F^2} \sqrt{\frac{ \tau\,\gamma_{\mathrm O}^{\,2} }{ (1-\varepsilon)c_{\mathrm O,eq}^{\,2} D_{\mathrm{chem}}\,a_s\,k_{\mathrm{chem}} } },
\end{equation}
and
\begin{equation}
    t_{\mathrm{chem}} = \frac{1-\varepsilon} {a_s k_{\mathrm{chem}}},
\end{equation}
where \(R_g\) is the universal gas constant, \(T\) is the operating temperature, \(F\) is the Faraday constant, \(D_{\mathrm{chem}}\) is the chemical oxygen diffusion coefficient, \(k_{\mathrm{chem}}\) is the chemical surface-exchange coefficient, \(c_{\mathrm O,eq}\) is the equilibrium oxygen concentration, \(\gamma_{\mathrm O}\) is the oxygen thermodynamic factor (characterizing the sensitivity of the oxygen chemical potential to changes in oxygen stoichiometry), \(\varepsilon\) is porosity, \(\tau\) is the solid-phase tortuosity factor, and \(a_s\) is the internal MIEC/gas surface area per unit electrode volume.

For a finite electrode thickness \(L\), the chemical impedance takes the finite-length form
\begin{equation}
    Z_{\mathrm{chem}}(\omega) = \frac{R_{\mathrm{chem}}} {\sqrt{1+i\omega t_{\mathrm{chem}}}} \tanh\!\left[ \frac{L}{\ell_\delta} \sqrt{1+i\omega t_{\mathrm{chem}}} \right],
\end{equation}
where
\begin{equation}
    \ell_\delta = \sqrt{ \frac{(1-\varepsilon)}{a_s\tau} \frac{D_{\mathrm{chem}}}{k_{\mathrm{chem}}}}
\end{equation}
is the characteristic penetration depth of the electrochemically active
region Note that here \(\omega\) denotes the \emph{angular} frequency of the applied electrochemical perturbation.
In the thick-electrode limit \(L\gg\ell_\delta\), the hyperbolic-tangent term approaches unity and the usual Gerischer-type ALS response is recovered; for thinner electrodes, the finite-thickness correction accounts for increasing utilization of the available electrode volume.

The electrode polarization resistance used for screening is obtained from the low-frequency limit,
\begin{equation}
    ASR \equiv R_p = \lim_{\omega\rightarrow0} \operatorname{Re} \left[Z_{\mathrm{chem}}(\omega)\right],
\end{equation}
or, when additional experimentally calibrated interfacial contributions are included, from the corresponding low-frequency non-ohmic impedance. 
This homogenized formulation is intentionally a screening-level model. 
Modern three-dimensional reconstructions show that local geometry and frequency-dependent effective tortuosity can produce corrections to the macro-homogeneous ALS approximation \cite{GoelEtAl2021}, while recent finite-element implementations retain the same surface-exchange/bulk-diffusion physics and explicitly predict ASR as a function of electrode thickness and morphology \cite{PanissetEtAl2024}. 
Such models provide a natural higher-fidelity validation layer for candidates retained by the present screening workflow.

\subsection{How the microscopic observables enter the MIEC model}
\label{sec:micro-inputs}
The embedding observables enter the MIEC continuum model through the defect thermodynamics and transport properties of the oxygen electrode.

\textbf{From vacancy and migration energies to chemical diffusivity.}
The vacancy-formation energy determines the equilibrium oxygen-vacancy population through defect thermodynamics. 
In the dilute-defect limit, 
\begin{equation}
    c_{\mathrm{vac}} = N_{\mathrm O} \exp\!\left( -\frac{\Delta G_{\mathrm{vac}}}{k_BT} \right),
\end{equation}
where \(N_{\mathrm O}\) is the density of available oxygen sites and \(\Delta G_{\mathrm{vac}}\) is the vacancy-formation free energy. 
Writing
\begin{equation}
    \Delta G_{\mathrm{vac}} = E_{\mathrm{vac}}^{\mathcal{I}} - T\Delta S_{\mathrm{vac}},
\end{equation}
shows explicitly how the embedding vacancy-formation energy, which itself depends on the oxygen chemical potential through Eq.~\eqref{eq:evac}, controls the equilibrium defect concentration together with entropic contribution \(\Delta S_{\mathrm{vac}}\). 
At fixed \(T\) and oxygen partial pressure, a lower \(E_{\mathrm{vac}}^{\mathcal{I}}\) therefore produces a larger equilibrium vacancy concentration.

The dilute-defect approximation assumes vacancy-independent formation energies, valid as long as vacancies are spaced far enough apart that their local electronic environments do not overlap. Because strongly correlated transition-metal $d$-states reorganize locally around each vacancy, this spacing requirement is typically satisfied at the low vacancy concentrations found in SOFC/SOEC operation. At higher concentrations, or if this electronic reorganization extends beyond the immediate coordination shell, the approximation may break down, representing a possible extension beyond the present screening-level treatment.

For aliovalently doped perovskites such as LSCF, charge neutrality couples the vacancy population to electronic defects and dopant concentration, so the full equilibrium concentration is obtained from the appropriate mass-action and charge-neutrality equations rather than from the dilute Boltzmann form alone.

The vacancy-formation energetics determine the equilibrium defect chemistry of the electrode. 
For a given temperature \(T\) and oxygen partial pressure \(p_{\mathrm O_2}\), the defect-chemistry model is solved subject to mass-action and charge-neutrality constraints to obtain the equilibrium oxygen-vacancy concentration
\begin{equation}
    c_{\mathrm{vac}} = c_{\mathrm{vac}}(T,p_{\mathrm O_2}).
\end{equation}
Repeating this calculation over a range of \(p_{\mathrm O_2}\) values at fixed temperature produces the equilibrium defect isotherm \(c_{\mathrm{vac}}(p_{\mathrm O_2})\). 
The corresponding concentration of occupied oxygen sites is
\begin{equation}
    c_{\mathrm O}(p_{\mathrm O_2}) = N_{\mathrm O}-c_{\mathrm{vac}}(p_{\mathrm O_2}).
\end{equation}
The thermodynamic factor is obtained from the \emph{slope} of this isotherm.
Specifically, the oxygen activity \(a_{\mathrm O}\) is related to the oxygen chemical potential through
\begin{equation}
    \mu_{\mathrm O} = \mu_{\mathrm O}^{0} + k_BT\ln a_{\mathrm O},
\end{equation}
where \(\mu_{\mathrm O}^{0}\) is the standard-state oxygen chemical potential.
For equilibrium with ideal molecular oxygen gas,
\begin{equation}
    a_{\mathrm O} = \left(\frac{p_{\mathrm O_2}}{p^0}\right)^{1/2},
\end{equation}
where \(p^0\) is the reference pressure corresponding to \(\mu_{\mathrm O}^{0}\),
and therefore
\begin{equation}
    \gamma_{\mathrm O} \equiv \frac{\partial\ln a_{\mathrm O}}{\partial\ln c_{\mathrm O}} = \frac{1}{2} \frac{\partial\ln p_{\mathrm O_2}} {\partial\ln c_{\mathrm O}}.
\end{equation}

In practice, after calculating \(c_{\mathrm{vac}}\) at several oxygen partial pressures, \(\gamma_{\mathrm{O}}\) can therefore be evaluated numerically, for two neighboring points for example, as
\begin{equation}
    \gamma_{\mathrm{O}} \simeq \frac{1}{2} \frac{ \ln\!\left(p_{\mathrm O_2}^{(2)}/p_{\mathrm O_2}^{(1)}\right)}{ \ln\!\left(c_{\mathrm O}^{(2)}/c_{\mathrm O}^{(1)}\right) },
    \qquad
    c_{\mathrm O}^{(j)} = N_{\mathrm O}-c_{\mathrm{vac}}^{(j)}.
\end{equation}
Thus, \(E_{\mathrm{vac}}^{\mathcal{I}}\) enters the defect-chemistry calculation that determines \(c_{\mathrm{vac}}(T,p_{\mathrm O_2})\), and the variation of this vacancy concentration with oxygen partial pressure determines \(\gamma_{\mathrm{O}}\).

Furthermore, oxygen-vacancy migration controls the microscopic vacancy diffusivity,
\begin{equation}
D_{\mathrm{vac}}
\propto
\exp\!\left(
-\frac{E_{\mathrm{mig}}^{\mathcal{I}}}{k_BT}
\right).
\end{equation}
The oxygen chemical diffusion coefficient entering the MIEC model is then obtained from the defect diffusivity multiplied by the normalized thermodynamic factor,
\begin{equation}
    D_{\mathrm{chem}} \simeq D_{\mathrm{vac}}\,\frac{c_\mathrm{vac}}{c_{\mathrm{O}}}\,\gamma_{\mathrm{O}}.
\end{equation}
Thus, \(E_{\mathrm{mig}}^{\mathcal{I}}\) controls the kinetic mobility, while \(E_{\mathrm{vac}}^{\mathcal{I}}\) influences the equilibrium defect concentration and therefore the thermodynamic factor.

\textbf{Chemical surface exchange coefficient.}
The second intrinsic kinetic input to the MIEC model is the chemical surface exchange coefficient \(k_{\mathrm{chem}}\). 
In the present work, \(k_{\mathrm{chem}}\) is treated as an experimentally calibrated parameter.
A future extension of the atomistic workflow could calculate the elementary surface reaction energetics required to predict this quantity directly. 
Its microscopic origin provides a possible additional application of correlated electronic-structure calculations. 
Oxygen surface exchange involves elementary steps such as adsorption, dissociation, charge transfer, and incorporation into oxygen vacancies.
For transition-metal oxide electrodes, the energetics of these steps can depend on the correlated transition-metal \(d\) states and oxygen \(p\) states.
A possible extension would therefore calculate the activation free energies of the relevant surface reactions using a correlated embedding method with a quantum impurity solver. 
Elementary reaction rates could then be estimated through transition-state theory. 
These rates would enter a surface microkinetic model from which an effective \(k_{\mathrm{chem}}\) could be obtained by linearizing the net oxygen-exchange flux around the operating state.
This extension is not included in the presented DFT+DMFT workflow, and it stands as an additional opportunity for quantum computing. 
Its potential quantum advantage would depend on identifying surface-reaction instances for which classical correlated electronic-structure methods remain computationally limiting.


\textbf{Electronic conductivity as model selection criterion.}
The ALS model assumes sufficiently rapid electronic transport such that the electronic potential drop across the electrode can be neglected. 
Thus, the embedding electronic conductivity \(\sigma_e^{\mathcal{I}}\) provides a consistency condition for the MIEC approximation. 
For an electrode of thickness \(L\), the electronic potential drop can be conservatively estimated by
\begin{equation}
    \Delta V_e \simeq \frac{j_e L}{\sigma_e^{\mathrm{eff}}},
\end{equation}
where \(j_e\) is the electronic current density and \(\sigma_e^{\mathrm{eff}}\) is the effective electronic conductivity (i.e.~ including porosity correction) in the through-thickness direction.
We require \(\Delta V_e\leq\eta_{e,\max}\), where \(\eta_{e,\max}\) is the maximum electronic voltage drop allowed by screening criteria. 
This gives
\begin{equation}
   \sigma_e^{\mathrm{eff}} \geq \sigma_{e,\min}^{\mathrm{eff}} = \frac{j_e L}{\eta_{e,\max}}. 
\end{equation}
Using a simple porous-medium approximation,
\begin{equation}
    \sigma_e^{\mathrm{eff}} \simeq \frac{1-\varepsilon}{\tau_e}\sigma_e^\mathcal{I},
\end{equation}
where \(\tau_e\) denotes electronic-path tortuosity, the corresponding intrinsic conductivity threshold is
\begin{equation}
    \sigma_{e,\min}^\mathcal{I} = \frac{\tau_e}{1-\varepsilon} \frac{j_e L}{\eta_{e,\max}}.
\end{equation}
For illustration, at \(j_e=0.25~\mathrm{A\,cm^{-2}}\), \(L=30~\mu\mathrm{m}\), and \(\eta_{e,\max}=5~\mathrm{mV}\), the required effective conductivity is \(0.15~\mathrm{S\,cm^{-1}}\).
If this criterion is satisfied by the embedding electronic conductivity, the electronically equipotential ALS approximation is retained. 
Otherwise, the candidate is evaluated using an extended MIEC transport model that explicitly couples ionic and electronic current conservation, chemical capacitance, and interfacial reaction kinetics \cite{BumbergerEtAl2024}.
The Costamagna framework \cite{Costamagna1998MicroModelling} is another alternative fallback for electrodes with distinct percolating ionic and electronic phases.

\textbf{Operational summary.}
The resulting hierarchy is therefore
\[
\left\{ E_{\mathrm{vac}}^{\mathcal{I}}, E_{\mathrm{mig}}^{\mathcal{I}}, \sigma_e^{\mathcal{I}} \right\} \longrightarrow \left\{ c_{\mathrm{vac}}, D_{\mathrm{chem}}, \text{electronic transport} \right\}, \]
which, together with
\[
\left\{ k_{\mathrm{chem}}, \gamma_{\mathrm O}, \varepsilon, \tau, a_s, L \right\},
\]
determines the electrode impedance and hence the predicted ASR. The quantum-computed observables therefore remain directly connected to the engineering screening metric, while experimentally calibrated kinetic and microstructural quantities complete the continuum model.

\subsection{Engineering targets}
The screening problem should be tied to explicit engineering targets. 
In SOFC service, projected DOE stack goals include \(>60\%\) efficiency, \(40{,}000\)~h lifetime, degradation below \(0.2\%\) per 1000 h, stack cost of \( 225\,\mathrm{USD}/\mathrm{kW}\), and system cost of \( 900 \,\mathrm{USD}/\mathrm{kW}\) \citep{DOE_SOFC_Program}. 
For high-temperature electrolysis, DOE targets include stack performance of \(1.2\,\mathrm{A/cm}^2\) at \(1.28\,\mathrm{V/cell}\), \(40{,}000\) h lifetime by 2026 and \(80{,}000\) h ultimate lifetime, stack cost of \(125 \,\mathrm{USD}/\mathrm{kW}\) by 2026, and system electricity consumption of \(36\,\mathrm{kWh/kg\ H_2}\) by 2026 \citep{DOE_HTE_Targets}. 
Motivated by these public targets, we take practical oxygen-electrode screening targets to be:
\[
500^\circ\mathrm{C} \le T_{\mathrm{op}} \le 700^\circ\mathrm{C},
\qquad
50{,}000\text{--}100{,}000\ \mathrm{h},
\]
and an electrode overpotential target of approximately \(50\,\mathrm{mV}\) at \(0.25\,\mathrm{A/cm}^2\), corresponding to an electrode ASR target of
\begin{equation}
    ASR_{\mathrm{target}} = \frac{0.05\ \mathrm{V}}{0.25\ \mathrm{A/cm}^2} = 0.20\ \Omega\,\mathrm{cm}^2.
\end{equation} 
We treat this ASR target as a design criterion for screening rather than as a universal industry standard.

Importantly, a favorable ASR is a necessary but not sufficient condition for
a viable oxygen-electrode material. 
The MIEC micromodel used here evaluates one part of the design problem: oxygen surface-exchange kinetics, bulk oxygen transport, defect thermodynamics, electrode thickness, porosity, tortuosity, and internal reactive surface area, with electronic transport treated explicitly when it is not sufficiently fast to be neglected. 
It does not by itself establish overall material or device suitability. We have listed in Section \ref{other-material-screening-criteria} other criteria that need to be taken into account in a comprehensive screening.

\subsection{Amplification of microscopic errors at the engineering level}
\label{sec:amplification}
The exponential dependence of oxygen transport on migration energetics means that relatively small errors in the microscopic activation energies can produce materially larger errors in predicted electrode resistance.
To illustrate this effect, consider the finite-thickness MIEC model introduced above.
The chemical diffusion coefficient has an Arrhenius dependence,
\begin{equation}
    D_{\mathrm{chem}} \propto \exp\!\left( -\frac{E_D}{k_BT} \right),
\end{equation}
where 
\begin{equation}
    E_D \simeq E_{\mathrm{mig}}^\mathcal{I} - k_B \frac{\partial \ln \left(\frac{c_\mathrm{vac}}{c_{\mathrm{O}}}\,\gamma_{\mathrm{O}} \right)} {\partial (1/T)},
\end{equation}
is the effective activation energy containing the microscopic contributions controlling oxygen mobility and defect thermodynamics. 

In the thick-electrode limit, \(L\gg\ell_\delta\), the ALS resistance scales as
\begin{equation}
    ASR \propto D_{\mathrm{chem}}^{-1/2} k_{\mathrm{chem}}^{-1/2}.
\end{equation}
Holding the surface-exchange coefficient fixed therefore gives
\begin{equation}
    \frac{\delta ASR}{ASR} \simeq \frac{1}{2} \frac{\delta E_D}{k_BT}.
\end{equation}

For illustration, suppose a migration energy \(E_{\mathrm{mig}}^\mathcal{I} =0.5~\mathrm{eV}\) at \(T=973~\mathrm{K}\), where \(k_BT\simeq0.084~\mathrm{eV}\), a one-percentage-point improvement in the relative accuracy of the migration barrier corresponds to
\begin{equation}
    \delta E_{\mathrm{mig}}^\mathcal{I} = 0.01\times0.5 = 0.005~\mathrm{eV}.
\end{equation}
The contribution of $E_{\mathrm{mig}}^\mathcal{I}$ to the uncertainty in the chemical diffusion coefficient is approximately
\begin{equation}
    \frac{\delta D_{\mathrm{chem}}}{D_{\mathrm{chem}}} \simeq \frac{0.005}{0.084} \simeq 6\%,
\end{equation}
and therefore produces approximately
\begin{equation}
    \frac{\delta ASR}{ASR} \simeq 3\%
\end{equation}
uncertainty in the predicted ASR in the thick-electrode limit.

The amplification depends on electrode thickness. In the thin-electrode limit, \(L\ll\ell_\delta\), the finite-thickness expression reduces approximately to
\begin{equation}
ASR\propto D_{\mathrm{chem}}^{-1},
\end{equation}
so that the same microscopic energy improvement produces an approximately \(6\%\) change in the predicted ASR. 
Thus the engineering value of improved microscopic accuracy is itself regime dependent: candidate electrodes whose performance is bulk-diffusion limited are more sensitive to migration-energy accuracy than those dominated by surface exchange \cite{PanissetEtAl2023,PanissetEtAl2024}.

The framework therefore provides a quantitative route for translating improved microscopic prediction accuracy into improved confidence in engineering-level material ranking.

\section{Market impact and business value}
We estimate the economic value of quantum-enabled materials discovery using a bottom-up approach that translates improvements in oxygen-electrode properties into specific economic consequences at the cell and system level.
We first define future SOEC and SOFC deployment scenarios and then estimate, for each installed gigawatt, the annual value associated with improvements in operating performance, system cost, and durability.
These per-unit benefits are multiplied by the relevant installed or newly deployed capacity and supplemented by the value of faster materials R\&D. SOEC and SOFC are treated separately because the same materials improvement creates value through
different mechanisms in electrolysis and power-generation operation.

\subsection{Value levers}
Our bottom-up market value model, estimating the yearly value for SOEC and SOFC uses four value levers:

\paragraph{Lower operating temperature.}
Reducing the oxygen-electrode polarization resistance, expressed here as ASR, allows acceptable performance at lower \(T\), which can reduce degradation, easing materials constraints and balance-of-plant requirements.

\paragraph{Higher efficiency.}
Lower ASR reduces voltage loss:
\begin{equation}
    \Delta V = j\,\Delta ASR
\end{equation}
where \(j\) is the operating current density of the device.
In SOEC mode this reduces electricity consumption per kilogram of hydrogen. In SOFC mode it increases electrical efficiency.

\paragraph{Longer lifetime.}
Durability also improves via reduced resistance, reducing replacement frequency and service cost.

\paragraph{Faster R\&D screening.}
Interface-aware quantum-enhanced embedding can eliminate poor electrode/electrolyte pairs early, reducing synthesis and testing cycles.

\subsection{SOEC global value pool}
Capacity for SOECs is measured in gigawatts (GW), capturing the rate at which electrical energy can be converted into hydrogen.
Improvements to the electrode materials lead to higher energy efficiency, thus lower energy requirements per volume of hydrogen produced. 
In this model, we estimate the total volume of energy saved for a fixed volume of hydrogen produced: combined with the price of electricity, we can calculate the value generated by the improved electrolytic cell design. 
IEA reports that installed electrolysis capacity for dedicated hydrogen production reached \(1.4\) GW by the end of 2023 and that announced projects could lead to \(230\text{--}520\) GW by 2030, although only around \(20\) GW had reached final investment decision or construction status at the time of reporting \citep{IEA_Electrolysers_2024,IEA_GlobalHydrogenReview2024}. 
We therefore use stylized SOEC deployment scenarios of \(10\), \(25\), and \(50\) GW by 2030, representing a modest share (5-20\%) of the broader electrolyser pipeline, making our model rather conservative.

For \(1\) GW of installed SOEC capacity, assuming an \(80\%\) capacity factor and \(40\,\mathrm{kWh/kg\ H_2}\) electricity consumption, annual hydrogen production is
\begin{equation}
    \frac{1{,}000{,}000 \times 8760 \times 0.8}{40}=175.2\times 10^6\ \mathrm{kg\ H_2/yr}.
\end{equation}
Assume a representative oxygen-electrode improvement of \(\Delta ASR = 0.05\,\Omega\mathrm{cm}^2\) at \(j=1\,\mathrm{A/cm}^2\), implying \(\Delta V=0.05\,\mathrm{V}\). Taking an experimentally reported baseline oxygen-electrode ASR of 0.32 \(\Sigma\)·cm² at 700°C for a LSCF–GDC composite \cite{BuxbaumEtAl2025}, the assumed improvement corresponds to a 15.6\% reduction in resistance. In the thick-electrode approximation of Section \ref{sec:amplification}, this could be achieved through an approximately 5.7\% reduction in a 0.5 eV migration barrier (i.e.\ 0.031 eV), holding other MIEC parameters fixed. Then
\begin{equation}
    \Delta \mathrm{kWh/kg\ H_2} \approx 26.6 \times \Delta V = 1.33\,\mathrm{kWh/kg\ H_2},
\end{equation}
and at \(40\,\mathrm{USD/MWh}\) electricity price the saving is \(0.053\,\mathrm{USD/kg\ H_2}\), or \(9.3\) M USD per GW-year.

For lower-temperature operation, we assume the improved oxygen electrode reduces effective system cost on new SOEC installations by \(5\%\). Using the Arrhenius dependence of electrode resistance derived in Section \ref{sec:amplification}, the illustrative 0.031 eV reduction in the effective energy from 0.5 eV would allow the operating temperature to decrease from 700°C to down to 640°C, under otherwise unchanged transport parameters.
The underlying rationale behind system-cost savings is that lower operating temperatures relax the thermal requirements imposed on stack and balance-of-plant components, enabling the use of lower-cost materials for components such as interconnects, seals, insulation, and piping, while potentially simplifying
system integration and thermal management \citep{AyersMarina2024}. 
Using DOE's \(500 \,\mathrm{USD}/\mathrm{kW}\) 2026 system target as a public anchor but retaining a stylized \(300\,\mathrm{USD/kW}\) working assumption for a future mature case, and assuming annual new installations equal \(20\%\) of installed capacity, this yields \(30\), \(75\), and \(150\) M USD/year in the three scenarios. 
For lifetime, we assume a stack-cost proxy of \(100\,\mathrm{USD/kW}\) and a barrier-enabled lifetime improvement from \(50{,}000\) h to \(75{,}000\) h. The same illustrative reduction in operating temperature from 700°C to 640°C can indeed slow thermally activated degradation mechanisms, including interfacial reactions and cation diffusion, providing a physical rationale. At \(80\%\) capacity factor (accounting for downtime and maintenance), this corresponds to \(7.1\) to \(10.7\) years, implying \(4.7\) M USD/GW-year replacement savings. 
For R\&D, we assume a global annual value of \(50\), \(100\), and \(200\) M USD/year. Combining these yields the SOEC value pool in Table~\ref{tab:soec}.

\begin{table}[t]
\centering
\caption{Stylized global SOEC value sizing.}
\label{tab:soec}
\begin{tabular}{lccc}
\toprule
Scenario & Conservative & Base & Upside \\
\midrule
Installed capacity (GW) & 10 & 25 & 50 \\
Efficiency value (M USD/yr) & 93 & 233 & 466 \\
Lower-temperature value (M USD/yr) & 30 & 75 & 150 \\
Lifetime value (M USD /yr) & 47 & 117 & 234 \\
R\&D screening value (M USD/yr) & 50 & 100 & 200 \\
\midrule
Total SOEC value (M USD/yr) & 220 & 525 & 1050 \\
\bottomrule
\end{tabular}
\end{table}

Thus the SOEC value pool is approximately \(0.2\text{--}1.1\) B USD/year, with a base case around \(0.5\) B USD/year.

\subsection{SOFC global value pool}
For SOFCs we use stylized deployment scenarios of \(5\), \(15\), and \(30\) GW by 2030. 
These values are scenarios rather than externally published capacity forecasts, anchored to the observed commercial scale and projected growth of stationary SOFC deployment. 
Bloom Energy, currently the largest commercial supplier of stationary SOFC systems, reports more than \(1.5~\mathrm{GW}\) of fuel-cell generation deployed across more than 1,200 installations globally \cite{BloomEnergyServer2026}. 
Recent contracted projects indicate that multi-gigawatt additions are already becoming commercially relevant: Bloom's agreement with Oracle alone provides for up to \(2.8~\mathrm{GW}\) of fuel-cell capacity, including an initial \(1.2~\mathrm{GW}\) already contracted and under deployment \cite{BloomOracle2026}. 
The stationary SOFC market is also projected to grow strongly, from approximately 3.2 billion USD in 2025 to 9.6 billion USD by 2030 (a compound annual growth rate of 24.4\%) \citep{BCC2025SOFCMarket}. Together, these observations make a future global installed base in the \(5\)--\(30~\mathrm{GW}\)  range a reasonable scenario set for an order-of-magnitude value analysis.
The range should nevertheless be interpreted as a sensitivity assumption rather than a forecast. Public DOE targets for SOFCs anchor costs and performance: \(>60\%\) efficiency, \(225 \,\mathrm{USD}/\mathrm{kW}\) stack cost, \(900 \,\mathrm{USD}/\mathrm{kW}\) system cost, and \(40{,}000\) h lifetime with low degradation \citep{DOE_SOFC_Program}. 

For a \(1\) GW SOFC fleet operating at \(80\%\) capacity factor, annual electricity output is
\begin{equation}
    E_{\mathrm{el}} = 1\times 8760 \times 0.8 = 7.008\ \mathrm{TWh/yr}.
\end{equation}
Assume a baseline electrical efficiency of \(60\%\), a baseline cell voltage \(V_{\mathrm{cell}}=0.8\,\mathrm{V}\), and again \(\Delta ASR =0.05\,\Omega\mathrm{cm}^2\) at \(j=1\,\mathrm{A/cm}^2\), implying \(\Delta V=0.05\,\mathrm{V}\). 
Approximating the relative efficiency gain by \(\Delta V/V_{\mathrm{cell}}=6.25\%\), the improved efficiency is \(63.75\%\). Then the fuel saved is
\begin{equation}
    \Delta E_{\mathrm{fuel}} = \frac{7.008}{0.60} - \frac{7.008}{0.6375} = 0.69\ \mathrm{TWh}_{\mathrm{fuel}}/\mathrm{GW\cdot yr}.
\end{equation}
At \(30\,\mathrm{USD/MWh_{fuel}}\), this is \(20.7\) M USD/GW-year. 
For stack-cost savings on new installations we use a \(5\%\) reduction applied to a \(225\,\mathrm{USD/kW}\) stack cost and \(20\%\) annual reinvestment rate. 
For lifetime we again assume an improvement from \(50{,}000\) h to \(75{,}000\) h, which implies \(10.5\) M USD/GW-year replacement savings at \(225\,\mathrm{USD/kW}\). 
For faster R\&D we assume \(25\), \(75\), and \(150\) M USD/year. Results are summarized in Table~\ref{tab:sofc}.

\begin{table}[t]
\centering
\caption{Stylized global SOFC value sizing.}
\label{tab:sofc}
\begin{tabular}{lccc}
\toprule
Scenario & Conservative & Base & Upside \\
\midrule
Installed capacity (GW) & 5 & 15 & 30 \\
Fuel-savings value (M USD/yr) & 104 & 311 & 622 \\
Stack-cost value (M USD/yr) & 11 & 34 & 68 \\
Lifetime value (M USD/yr) & 53 & 158 & 315 \\
R\&D screening value (M USD/yr) & 25 & 75 & 150 \\
\midrule
Total SOFC value (M USD /yr) & 193 & 578 & 1155 \\
\bottomrule
\end{tabular}
\end{table}

Thus the SOFC value pool is approximately \(0.2\text{--}1.2\) B USD/year, with a base case around \(0.6\) B USD/year.

\subsection{Combined global value pool from bottom-up model}
Our bottom-up model yields a combined SOEC and SOFC annual gross value (i.e.~value creation) pool of approximately
\[
0.41,\quad 1.11,\quad 2.21\ \mathrm{B \; USD/yr}
\]
for the conservative, base, and upside cases. 
The gross value is enabled by better oxygen-electrode materials and interfaces, reflecting the fact that quantum computing improves a high-impact subproblem of solide-oxide cell design.

\subsection{Top-down validation of the market value estimates}
As a top-down validation, we compare the bottom-up value pool estimated above with broader estimates of the economic value of fault-tolerant quantum computing in materials design. 
BCG analysis estimates that, at maturity, quantum-enabled materials design could generate approximately 27-55~billion USD of annual economic value across industries \cite{BCG2019QuantumValue}. 
A more fine-grained analysis (unpublished) of internal data underlying these estimates reveal that approximately 35.3~billion USD of annual value are directly relevant to our use case, given the presently considered value levers. 
This comprises 11.9~billion USD per year of cost savings, including approximately 9.3~billion USD in R\&D expenditure, together with 16.7~billion USD per year from net-new materials and 6.7~billion USD per year from improvements to existing materials. 
More recent published BCG work similarly identifies chemistry and materials simulation among the high-value applications expected to become accessible to early fault-tolerant quantum computers \cite{BCG2026QuantumValue}.

Against this cross-industry value pool, the combined SOFC/SOEC value estimated in our bottom-up model---~0.41~billion, ~1.11~billion, and ~2.21~billion USD per year in the conservative, base, and upside scenarios, respectively---corresponds approximately to
\begin{equation}
    \frac{0.41}{35.3}\simeq1.2\%,
    \qquad
    \frac{1.11}{35.3}\simeq3.1\%,
    \qquad
    \frac{2.21}{35.3}\simeq6.3\%
\end{equation}
of the total materials-design value pool. 
The different scenarios therefore imply that quantum-enabled materials design for the specific solid-oxide fuel-cell and electrolysis technologies considered here could account for approximately \(1-6\%\) of the total economic value attributed to quantum-enabled materials design across all materials and industries. 
Given the narrow scope of the present application relative to the full materials-design opportunity---which also includes batteries, catalysts, chemicals, structural materials, semiconductors, and other applications---this share provides a first directionally consistent top-down validation of the bottom-up estimate.

Furthermore, because the economic value estimated here is intended to materialize in the fault-tolerant quantum-computing era, a comparison with the projected rather than current market for the underlying technology provides an additional top-down consistency check. 
Grand View Research projects the global market for stationary solid-oxide fuel cells (SOFCs) to reach approximately 10.1~billion USD per year by 2033 \cite{GVR2026StationarySOFC}. 
This is the closest available market benchmark for the SOFC application considered here, as it specifically covers stationary electricity generation using the same solid-oxide technology. 

Our base-case estimate of approximately ~0.58~billion USD per year of value creation from improved SOFC oxygen-electrode materials therefore represents approximately
\begin{equation}
\frac{0.58}{10.1}\simeq 5.7\%
\end{equation}
of the projected stationary SOFC market. 
The conservative and upside estimates of ~0.19~billion and ~1.16~billion USD correspond to approximately \(1.9\%\) and \(11.5\%\), respectively. 
These ratios appear directionally consistent with the scope of the intervention: the proposed computation does not create the complete SOFC system, but targets oxygen-electrode materials whose properties directly affect cell efficiency, power density, degradation, and stack lifetime. 
As a broader independent benchmark, BCG estimates that equipment for hydrogen-based power and heat generation could represent an 8--10~billion USD annual market by 2050, including both stationary fuel cells and hydrogen combustion technologies \cite{BCG2021GreenHydrogen}. 
The similar order of magnitude obtained from these independent market projections provides an additional top-down consistency check on the bottom-up value estimate.

\section{Discussion}
Two architectural points are worth emphasizing.
First, the role of the electrolyte is not to add extra observables.
Rather, it changes the atomistic structure and therefore the DFT-derived Hamiltonian from which the local impurity models are built.
The core observables must therefore be computed in the interfacial environment.
Second, the impurity solver remains local even in the interface case; the quantum computer is solving the Anderson impurity model, not the full heterogeneous device. This is exactly why the application is plausible at all \citep{BauerEtAl2016}.

The remaining scientific risk lies in model calibration, in the accuracy of the structural and electronic approximations, and, specific to the quantum step, in whether a quantum impurity solver is needed and affordable at the required accuracy (Section~\ref{sec:open}).
Three of these deserve particular attention. The first is the maturity of total energies beyond DFT+DMFT: to our knowledge, GW-based embeddings have not yet delivered defect energetics in large relaxed supercells of transition-metal oxides (Section~\ref{sec:beyond-dmft}). The second is the use of DFT geometries with beyond-DFT energies (Section~\ref{sec:geometry}). The third is the classical cost of the GW step in Tier T1. If this step dominates, accelerating the impurity solver gives only a limited reduction in total cost (Section~\ref{sec:open}).
The choice of correlated subspace, determination of \(U\) and \(J\), stability of total-energy differences in the chosen embedding, and bath discretization needed for reliable barrier estimates are all nontrivial.
In addition, the baseline workflow inherits the atomic geometries produced by DFT relaxation; where correlation effects materially change the forces or transition-state geometry, a DFT-relaxed structure may itself become the dominant source of error.
DFT+DMFT force calculations and structural optimization provide a route to remove this assumption, although at substantially greater computational cost \cite{HauleBirol2015,HaulePascut2016,PlekhanovEtAl2021}.
These are expert-facing implementation and validation questions that must ultimately be benchmarked against experiment.

A natural first step is a validation benchmark before the interface calculations.
The GW-based embedding is run on bulk LSCF at increasing active-space size, with DFT+\(U\) as a reference, and its \(\mu_{\mathrm O}(\delta)\) and \(E_{\mathrm{vac}}\) are compared with thermogravimetry \citep{Hashimoto2010LSCF} before the interface is treated.
This benchmark also fixes the active-space size needed for converged energy differences before the interface calculations.

\subsection{From a single use case to a library of quantum-relevant
materials problems}
The SOFC/SOEC example should be viewed as one instance of a broader research program rather than as an isolated quantum-computing application.
The longer-term objective is to construct a curated library of industrial materials-design problems in which strong electronic correlations create a well-defined limitation for conventional computational screening and in which DMFT maps that limitation onto quantum-computable impurity models. 
Candidate instances could span transition-metal oxides for batteries and catalysis, correlated electrocatalysts, magnetic and multiferroic materials, rare-earth compounds, and other systems containing partially filled localized \(d\) or \(f\) shells \cite{IsaacsMarianetti2020,BiFeO3DMFT2018,XieZhang2022}.
The relevant unit of the library would not simply be a material composition, but a complete computational instance: crystal or interface geometry, correlated subspace, interaction parameters, impurity Hamiltonian, target temperature and accuracy, classically computed baseline, experimentally measurable target observable, and the engineering decision affected by that observable, together with the active-space size that each observable needs.

Such a library would serve three purposes.
First, it would allow classical methods---including DFT, DFT+\(U\), and state-of-the-art classical DMFT solvers---to be benchmarked systematically against experiment, identifying where additional many-body accuracy has genuine decision value. 
Second, it would provide concrete Anderson impurity Hamiltonians on which competing quantum algorithms and fault-tolerant architectures could be compared using common accuracy and resource metrics. 
Third, it would connect quantum resource requirements to industrial value: for each instance, one could ask not only how many logical qubits and gates are required, but whether the resulting improvement in predictive accuracy changes a material-selection decision sufficiently to justify the computation. 
The present LSCF/GDC workflow provides a template for constructing the first such instances.

\subsection{Open problems for the quantum step}
\label{sec:open}
Quantum impurity solvers offer a potential route to extending the correlated active spaces accessible to materials screening. Establishing their practical advantage requires resolving four questions concerning the required impurity model, quantum execution costs, observable-level accuracy, and the complete embedding workload.

\begin{itemize}
    \item \textbf{Required impurity model and classical baseline.}
    Establish convergence of \(E_{\mathrm{vac}}\) and \(E_{\mathrm{mig}}\) with correlated active-space size and bath discretization, and benchmark the best applicable classical impurity solvers at the target temperature.
    Bulk LSCF provides an initial benchmark, followed by tests on the relevant interface configurations.
    For GW+DMFT, the comparison must include classical solvers for dynamical monopole screening with static exchange.

    \item \textbf{Quantum preparation and simulation cost.}
    Determine the cost of thermal-state preparation and Hamiltonian simulation for the converged impurity models.
    This includes the block-encoding normalization and query implementation, or the corresponding costs of alternative simulation methods.
    For dynamical screening, convergence must also be established with respect to the number of bosonic modes and their occupation cutoffs.

    \item \textbf{Precision and finite-temperature energetics.}
    Quantify how errors in measured correlators, bath fitting and incomplete self-consistency propagate to the target energy differences and candidate rankings.
    Controlled-noise classical calculations can establish the required measurement precision.
    The chosen embedding free-energy functional must also be implemented and evaluated consistently for these models.

    \item \textbf{End-to-end resources.}
    Combine the cost per impurity solve with the number of geometries, correlated regions and self-consistency iterations.
    Include the classical GW and embedding workload, which may limit the total speedup, and translate the logical circuit costs into physical qubits and runtime for a specified fault-tolerant architecture.
\end{itemize}

Classical polynomial-time guarantees for equilibrium properties of constant-size impurities coupled to fermionic baths \citep{JiangEtAl2026,ArunachalamEtAl2026} provide important context, but do not establish practical efficiency for the enlarged active spaces and screening models considered here.
A formal complexity separation is not required to demonstrate a practical advantage.
The decisive comparison is the total cost of obtaining predictions at the accuracy needed for material screening.

\section{Conclusion}
We have presented an end-to-end application architecture for quantum computing in industrial materials science, using strongly correlated oxygen-electrode screening for SOFCs and SOECs as the concrete use case. 
The architecture begins with an interface-aware materials problem, continues through a DFT screen and a GW-based embedding (GW+DMFT or SEET) whose local impurity problems are solved on a quantum computer, maps the resulting observables into an electrode-resistance model, and ends with deployment-scale business value. 

The correlated calculations use GW-based embedding (GW+DMFT or SEET), applied with the quantum impurity solver to a DFT-screened shortlist.
We propose to select between these options through a bulk LSCF benchmark against experimental thermodynamic data and classical impurity calculations. 
The quantum computer solves impurity problems of the target class defined in Section~\ref{sec:quantum-algos}, with the correlated active space enlarged until total-energy differences converge, if that active space lies beyond the best classical impurity solvers, which the bulk benchmark and the comparison with the best classical impurity solvers test.

The result is a practical translation layer between four communities that rarely write in the same language: materials scientists, condensed-matter physicists, quantum-computing researchers, and industrial strategists.
In that sense, this paper's main contribution is neither purely scientific (and computational) nor purely economic.
It is a proposal for a coherent application stack, together with the open problems that decide whether its quantum part pays off.

\bibliographystyle{unsrtnat}
\bibliography{references}

@article{ParkEtAl2014,
  author  = {Park, Hyowon and Millis, Andrew J. and Marianetti, Chris A.},
  title   = {Computing Total Energies in Complex Materials Using
             Charge Self-Consistent DFT+DMFT},
  journal = {Physical Review B},
  volume  = {90},
  number  = {23},
  pages   = {235103},
  year    = {2014},
  doi     = {10.1103/PhysRevB.90.235103}
}

@book{Chase1998JANAF,
  author    = {Chase, Malcolm W., Jr.},
  title     = {NIST-JANAF Thermochemical Tables},
  edition   = {4},
  series    = {Journal of Physical and Chemical Reference Data,
               Monograph},
  number    = {9},
  publisher = {American Institute of Physics},
  year      = {1998},
  isbn      = {978-1-56396-831-0},
  url       = {https://janaf.nist.gov/}
}

@article{IsaacsMarianetti2020,
  author  = {Isaacs, Eric B. and Marianetti, Chris A.},
  title   = {Compositional Phase Stability of Correlated Electron
             Materials within DFT+DMFT},
  journal = {Physical Review B},
  volume  = {102},
  number  = {4},
  pages   = {045146},
  year    = {2020},
  doi     = {10.1103/PhysRevB.102.045146}
}

@article{XieZhang2022,
  author  = {Xie, Ruiwen and Zhang, Hongbin},
  title   = {Mixed Valence Nature of the Ce 4f State in CeCo5
             Based on Spin-Polarized DFT+DMFT Calculations},
  journal = {Physical Review B},
  volume  = {106},
  pages   = {224411},
  year    = {2022},
  doi     = {10.1103/PhysRevB.106.224411}
}

@article{AdlerLaneSteele1996,
  author  = {Adler, Stuart B. and Lane, J. A. and Steele, B. C. H.},
  title   = {Electrode Kinetics of Porous Mixed-Conducting Oxygen Electrodes},
  journal = {Journal of The Electrochemical Society},
  volume  = {143},
  number  = {11},
  pages   = {3554--3564},
  year    = {1996},
  doi     = {10.1149/1.1837252}
}

@article{Adler1998,
  author  = {Adler, Stuart B.},
  title   = {Mechanism and Kinetics of Oxygen Reduction on Porous
             La1-xSrxCoO3-delta Electrodes},
  journal = {Solid State Ionics},
  volume  = {111},
  pages   = {125--134},
  year    = {1998},
  doi     = {10.1016/S0167-2738(98)00179-9}
}

@article{PanissetEtAl2023,
  author  = {Panisset, Silv{\`e}re and Burriel, M{\'o}nica
             and Laurencin, J{\'e}r{\^o}me and Jauffres, David},
  title   = {Modelling of Solid Oxide Cell Oxygen Electrodes},
  journal = {Journal of Physics: Energy},
  volume  = {5},
  number  = {2},
  pages   = {022003},
  year    = {2023},
  doi     = {10.1088/2515-7655/acc5b1}
}

@article{MortensenEtAl2014,
  author  = {Mortensen, Jakob E. and S{\o}gaard, Martin and Jacobsen, Torben},
  title   = {Analytical, 1-Dimensional Impedance Model of a Composite
             Solid Oxide Fuel Cell Cathode},
  journal = {Journal of The Electrochemical Society},
  volume  = {161},
  number  = {3},
  pages   = {F161--F175},
  year    = {2014},
  doi     = {10.1149/2.077401jes}
}

@article{GoelEtAl2021,
  author  = {Goel, Vishwas and Cox, Dalton and Barnett, Scott A.
             and Thornton, Katsuyo},
  title   = {Simulation of the Electrochemical Impedance in a
             Three-Dimensional, Complex Microstructure of Solid Oxide
             Fuel Cell Cathode and Its Application in the
             Microstructure Characterization},
  journal = {Frontiers in Chemistry},
  volume  = {9},
  pages   = {627699},
  year    = {2021},
  doi     = {10.3389/fchem.2021.627699}
}

@article{PanissetEtAl2024,
  author  = {Panisset, Silv{\`e}re and Riaz, A. and Stangl, A.
             and Burriel, M{\'o}nica and Jauffr{\`e}s, David},
  title   = {Contribution of Numerical Modelling to Design Oxygen Electrode
             for Micro-Solid Oxide Cells: A Case Study of High-Performance
             Nano-Columnar La2NiO4 Thin Films},
  journal = {Journal of Power Sources},
  volume  = {593},
  pages   = {233951},
  year    = {2024},
  doi     = {10.1016/j.jpowsour.2023.233951}
}

@article{ZurMcGill1984,
  author  = {Zur, A. and McGill, T. C.},
  title   = {Lattice Match: An Application to Heteroepitaxy},
  journal = {Journal of Applied Physics},
  volume  = {55},
  number  = {2},
  pages   = {378--386},
  year    = {1984},
  doi     = {10.1063/1.333084}
}

@article{BiFeO3DMFT2018,
  author  = {Paul, Souvik and Iu{\c{s}}an, Diana
             and Thunstr{\"o}m, Patrik and Kvashnin, Yaroslav O.
             and Hellsvik, Johan and Pereiro, Manuel
             and Delin, Anna and Knut, Ronny
             and Phuyal, Dibya and Lindblad, Andreas
             and Karis, Olof and Sanyal, Biplab
             and Eriksson, Olle},
  title   = {Investigation of the Spectral Properties and Magnetism
             of {BiFeO3} by Dynamical Mean-Field Theory},
  journal = {Physical Review B},
  volume  = {97},
  number  = {12},
  pages   = {125120},
  year    = {2018},
  doi     = {10.1103/PhysRevB.97.125120}
}

@article{BumbergerEtAl2024,
  author  = {Bumberger, Andreas E. and Nenning, Andreas
             and Fleig, Juergen},
  title   = {Transmission Line Revisited -- The Impedance
             of Mixed Ionic and Electronic Conductors},
  journal = {Physical Chemistry Chemical Physics},
  volume  = {26},
  pages   = {15068--15089},
  year    = {2024},
  doi     = {10.1039/D4CP00975D}
}

@article{Held2007ElectronicStructure,
  title = {Electronic Structure Calculations Using Dynamical Mean Field Theory},
  author = {Held, Karsten},
  journal = {Advances in Physics},
  volume = {56},
  number = {6},
  pages = {829--926},
  year = {2007},
  doi = {10.1080/00018730701619647}
}

@article{HauleBirol2015,
  author  = {Haule, Kristjan and Birol, Turan},
  title   = {Free Energy from Stationary Implementation of the
             DFT+DMFT Functional},
  journal = {Physical Review Letters},
  volume  = {115},
  pages   = {256402},
  year    = {2015},
  doi     = {10.1103/PhysRevLett.115.256402}
}

@article{HaulePascut2016,
  author  = {Haule, Kristjan and Pascut, Gheorghe L.},
  title   = {Forces for Structural Optimizations in Correlated Materials
             within a DFT+Embedded DMFT Functional Approach},
  journal = {Physical Review B},
  volume  = {94},
  pages   = {195146},
  year    = {2016},
  doi     = {10.1103/PhysRevB.94.195146}
}

@article{PlekhanovEtAl2021,
  author  = {Plekhanov, Evgeny and Bonini, Nicola and Weber, Cedric},
  title   = {Calculating Dynamical Mean-Field Theory Forces in
             Ab Initio Ultrasoft Pseudopotential Formalism},
  journal = {Physical Review B},
  volume  = {104},
  pages   = {235131},
  year    = {2021},
  doi     = {10.1103/PhysRevB.104.235131}
}

@article{OBrienEtAl2019Derivatives,
  author  = {O'Brien, Thomas E. and Senjean, Bruno and Sagastizabal, Ramiro
             and Bonet-Monroig, Xavier and Dutkiewicz, Alicja
             and Buda, Francesco and DiCarlo, Leonardo and Visscher, Lucas},
  title   = {Calculating Energy Derivatives for Quantum Chemistry
             on a Quantum Computer},
  journal = {npj Quantum Information},
  volume  = {5},
  pages   = {113},
  year    = {2019},
  doi     = {10.1038/s41534-019-0213-4}
}

@article{OBrienEtAl2022Forces,
  author  = {O'Brien, Thomas E. and Streif, Michael and Rubin, Nicholas C.
             and Santagati, Raffaele and Su, Yuan and Huggins, William J.
             and Goings, Joshua J. and Moll, Nikolaj and Kyoseva, Elica
             and Degroote, Matthias and Tautermann, Christofer S.
             and Lee, Joonho and Berry, Dominic W. and Wiebe, Nathan
             and Babbush, Ryan},
  title   = {Efficient Quantum Computation of Molecular Forces and
             Other Energy Gradients},
  journal = {Physical Review Research},
  volume  = {4},
  pages   = {043210},
  year    = {2022},
  doi     = {10.1103/PhysRevResearch.4.043210}
}

@misc{JiangEtAl2026,
  author        = {Jiang, Jiaqing and Ju, Nathan
                   and Parekh, Ojas and Rayudu, Chaithanya
                   and Zhao, Andrew},
  title         = {Polynomial-Time Classical and Quantum
                   Simulation of Quantum Impurity Models},
  year          = {2026},
  eprint        = {2610.02167},
  archivePrefix = {arXiv},
  primaryClass  = {quant-ph},
  doi           = {10.48550/arXiv.2610.02167},
  url           = {https://arxiv.org/abs/2610.02167}
}

@article{Costamagna1998MicroModelling,
  title = {Micro-Modelling of Solid Oxide Fuel Cell Electrodes},
  author = {Costamagna, Paola and Costa, Paolo and Antonucci, Vincenzo},
  journal = {Electrochimica Acta},
  volume = {43},
  number = {3--4},
  pages = {375--394},
  year = {1998},
  doi = {10.1016/S0013-4686(97)00063-7}
}

@article{Yang2004ReactionKineticsLSMYSZ,
  title = {Reaction Kinetics and Mechanisms between {La$_{0.65}$Sr$_{0.30}$MnO$_3$} and 8 mol\% Yttria-Stabilized Zirconia},
  author = {Yang, Chih-Chung and Wei, Wen-Cheng J. and Roosen, Andreas},
  journal = {Journal of the American Ceramic Society},
  volume = {87},
  number = {6},
  pages = {1110--1116},
  year = {2004},
  doi = {10.1111/j.1551-2916.2004.01110.x}
}

@article{He2023CriticalReview,
  title = {A Critical Review of Key Materials and Issues in Solid Oxide Cells},
  author = {He, Shuai and Zou, Yuanfeng and Chen, Kongfa and Jiang, San P.},
  journal = {Interdisciplinary Materials},
  volume = {2},
  number = {2},
  pages = {197--229},
  year = {2023},
  doi = {10.1002/idm2.12068}
}

@article{NisarEtAl2024,
  author  = {Nisar, Jamila and Kaur, Gurpreet and Giddey, Sarbjit
             and Bhargava, Suresh and Jones, Lathe},
  title   = {Cathode Materials for Intermediate Temperature Solid Oxide Fuel Cells},
  journal = {Fuels},
  volume  = {5},
  number  = {4},
  pages   = {805--824},
  year    = {2024},
  doi     = {10.3390/fuels5040045}
}

@article{Takahashi2010RuddlesdenPopper,
  title = {Electrode Properties of the Ruddlesden--Popper Series, {La}$_{n+1}${Ni}$_n${O}$_{3n+1}$ (\(n=1,2,3\)), as Intermediate-Temperature Solid Oxide Fuel Cells},
  author = {Takahashi, Suguru and Nishimoto, Shunsuke and Matsuda, Motohide and Miyake, Michihiro},
  journal = {Journal of the American Ceramic Society},
  volume = {93},
  number = {8},
  pages = {2329--2333},
  year = {2010},
  doi = {10.1111/j.1551-2916.2010.03743.x}
}

@article{Hu2022InterfacialNanocomposite,
  title = {Enhanced Long-Term Cathode Stability by Tuning Interfacial Nanocomposite for Intermediate Temperature Solid Oxide Fuel Cells},
  author = {Hu, D. and others},
  journal = {Advanced Materials Interfaces},
  volume = {9},
  number = {14},
  pages = {2102131},
  year = {2022},
  doi = {10.1002/admi.202102131}
}

@article{Yang2024ReversibleSOCs,
  title = {Recent Development in Reversible Solid Oxide Fuel Cells: Theory, Integration and Prospective},
  author = {Yang, Yiping and Lei, Jinyong and Huang, Xurui and Liao, Zihao and Liu, Yang and Tu, Zhengkai},
  journal = {ChemElectroChem},
  volume = {11},
  number = {2},
  pages = {e202300593},
  year = {2024},
  doi = {10.1002/celc.202300593}
}

@misc{DOE_SOFC_Program,
  author = {{U.S. Department of Energy}},
  title = {Solid Oxide Fuel Cells},
  year = {2026},
  howpublished = {\url{https://www.energy.gov/hgeo/solid-oxide-fuel-cells}},
  note = {Accessed 2026-07-30}
}

@misc{DOE_HTE_Targets,
  author = {{U.S. Department of Energy}},
  title = {Technical Targets for High Temperature Electrolysis},
  year = {2026},
  howpublished = {\url{https://www.energy.gov/cmei/fuels/technical-targets-high-temperature-electrolysis}},
  note = {Accessed 2026-07-30}
}

@misc{IEA_Electrolysers_2024,
  author = {{International Energy Agency}},
  title = {Electrolysers},
  year = {2024},
  howpublished = {\url{https://www.iea.org/energy-system/low-emissions-fuels/electrolysers}},
  note = {Accessed 2026-07-30}
}

@misc{IEA_GlobalHydrogenReview2024,
  author = {{International Energy Agency}},
  title = {Global Hydrogen Review 2024},
  year = {2024},
  howpublished = {\url{https://www.iea.org/reports/global-hydrogen-review-2024}},
  note = {Accessed 2026-07-30}
}

@article{LiEtAl2011LSCFGDC,
  author  = {Li, Zhi-Peng and Mori, Toshiyuki and
             Auchterlonie, Graeme John and Zou, Jin and Drennan, John},
  title   = {Mutual Diffusion Occurring at the Interface between
             {La0.6Sr0.4Co0.8Fe0.2O3} Cathode and Gd-Doped Ceria
             Electrolyte during {IT-SOFC} Cell Preparation},
  journal = {ACS Applied Materials \& Interfaces},
  volume  = {3},
  number  = {7},
  pages   = {2772--2778},
  year    = {2011},
  doi     = {10.1021/am2005543}
}

@article{LeeEtAl2016LSCFHeterostructure,
  author  = {Lee, Dongkyu and Lee, Yueh-Lin and Wang, Xiao Renshaw
             and Morgan, Dane and Shao-Horn, Yang},
  title   = {Enhancement of Oxygen Surface Exchange on Epitaxial
             {La0.6Sr0.4Co0.2Fe0.8O3-delta} Thin Films Using
             Advanced Heterostructured Oxide Interface Engineering},
  journal = {MRS Communications},
  volume  = {6},
  number  = {3},
  pages   = {204--209},
  year    = {2016},
  doi     = {10.1557/mrc.2016.28}
}

@article{AsakuraEtAl2026LSCFGDC,
  author  = {Asakura, Daisuke and Ishiyama, Tomohiro and Hosono, Eiji
             and Develos-Bagarinao, Katherine and Yamaji, Katsuhiko
             and Kobayashi, Masaki and Kitamura, Miho and Horiba, Koji
             and Kishimoto, Haruo},
  title   = {Investigation of the Interface Structure and Electronic State
             of Nanocomposite
             {La0.6Sr0.4Co0.2Fe0.8O3-delta} and
             {Ce0.9Gd0.1O2-delta} Electrodes for Solid Oxide Fuel Cells},
  journal = {Journal of Materials Chemistry A},
  pages= {8857--8865},
  volume  = {14},
  year    = {2026},
  doi     = {10.1039/D5TA05378A}
}

@article{HardyEtAl2012LSCF,
  author  = {Hardy, John S. and Templeton, Jared W. and Edwards, Danny J.
             and Lu, Zigui and Stevenson, Jeffry W.},
  title   = {Lattice Expansion of {LSCF-6428} Cathodes Measured by
             In-Situ {XRD} during {SOFC} Operation},
  journal = {Journal of Power Sources},
  volume  = {198},
  pages   = {76--82},
  year    = {2012},
  doi     = {10.1016/j.jpowsour.2011.09.099}
}

@article{LeeEtAl2017GDC,
  author  = {Lee, Ho Chang and Lee, Jung A. and Lee, Joon Hyung
             and Heo, Young Woo and Kim, Jeong Joo},
  title   = {Ionic Conductivity and Relaxations of In-Doped
             {GDC} (Gadolinium-Doped Ceria) Ceramics},
  journal = {Ceramics International},
  volume  = {43},
  number  = {15},
  pages   = {11792--11798},
  year    = {2017},
  doi     = {10.1016/j.ceramint.2017.06.018}
}

@article{BCG2019QuantumValue,
  author  = {Langione, Matt and Tillemann-Dick, Corban and Kumar, Amit
             and Taneja, Vikas},
  title   = {Where Will Quantum Computers Create Value---and When?},
  journal = {Boston Consulting Group},
  year    = {2019},
  month   = {May},
  note    = {Published May 13, 2019},
  url     = {https://www.bcg.com/publications/2019/quantum-computers-create-value-when}
}

@article{BCG2026QuantumValue,
  author  = {Langione, Matt and Park, Hanl and Bobier, Jean-Fran{\c{c}}ois
             and Henderson, Brad and Johnson, Harley T. and Iyer, Sesh
             and Cui, Zheng},
  title   = {Quantum Is Getting Real. CEOs Need to Shape Where It Creates Value},
  journal = {Boston Consulting Group},
  year    = {2026},
  month   = {June},
  note    = {Published June 4, 2026},
  url     = {https://www.bcg.com/publications/2026/ceos-need-to-shape-where-quantum-creates-value}
}

@misc{GVR2026StationarySOFC,
  author       = {{Grand View Research}},
  title        = {Stationary Solid Oxide Fuel Cell Market Statistics},
  year         = {2026},
  note         = {Global stationary SOFC market projected to reach
                  USD 10.13 billion by 2033},
  url          = {https://www.grandviewresearch.com/horizon/statistics/solid-oxide-fuel-cell-market/application/stationary/global}
}

@article{BCG2021GreenHydrogen,
  author  = {Ludwig, Max and L{\"u}ers, Martin and Lorenz, Markus
          and Hegnsholt, Esben and Kim, Minjee and Pieper, Cornelius
          and Meidert, Katharina},
  title   = {The Green Tech Opportunity in Hydrogen},
  journal = {Boston Consulting Group},
  year    = {2021},
  url     = {https://www.bcg.com/publications/2021/capturing-value-in-the-low-carbon-hydrogen-market}
}

@misc{BloomEnergyServer2026,
  author       = {{Bloom Energy}},
  title        = {Bloom Energy Server Data Sheet},
  year         = {2026},
  note         = {Bloom Energy reports more than 1.5 GW deployed across
                  more than 1,200 sites globally},
  url          = {https://www.bloomenergy.com/resource/bloom-energy-server/}
}

@article{MottaEtAl2020,
  author  = {Motta, Mario and Sun, Chong and Tan, Adrian T. K.
             and O'Rourke, Matthew J. and Ye, Erika and Minnich, Austin J.
             and Brandao, Fernando G. S. L. and Chan, Garnet Kin-Lic},
  title   = {Determining Eigenstates and Thermal States on a Quantum Computer
             Using Quantum Imaginary Time Evolution},
  journal = {Nature Physics},
  volume  = {16},
  pages   = {205--210},
  year    = {2020},
  doi     = {10.1038/s41567-019-0704-4}
}

@misc{DOE2020SOFCApplications,
  author = {{U.S. Department of Energy}},
  title  = {Project Selections for FOA 2300: Small-Scale Solid Oxide
            Fuel Cell Systems and Hybrid Electrolyzer Technology Development},
  year   = {2020},
  url    = {https://www.energy.gov/hgeo/articles/project-selections-foa-2300}
}

@misc{DOE2024SolidOxide,
  author = {{U.S. Department of Energy}},
  title  = {DOE Announces \$4 Million to Advance Solid Oxide Fuel Cell
            Technology for Clean, Low-Cost Hydrogen Production},
  year   = {2024},
  url    = {https://www.energy.gov/hgeo/articles/doe-announces-4-million-advance-solid-oxide-fuel-cell-technology-clean-low-cost}
}

@book{InternationalTables2016,
  editor    = {Aroyo, Mois I.},
  title     = {International Tables for Crystallography, Volume A:
               Space-Group Symmetry},
  edition   = {6},
  publisher = {International Union of Crystallography},
  year      = {2016},
  isbn      = {978-0-470-97423-0},
  doi       = {10.1107/97809553602060000114},
  url       = {https://doi.org/10.1107/97809553602060000114}
}

@article{AyersMarina2024,
  author  = {Ayers, Katherine E. and Marina, Olga A.},
  title   = {State of the Art in Low-Temperature and High-Temperature Electrolysis},
  journal = {MRS Bulletin},
  volume  = {49},
  number  = {12},
  pages   = {1226--1234},
  year    = {2024},
  doi     = {10.1557/s43577-024-00806-6},
  url     = {https://link.springer.com/article/10.1557/s43577-024-00806-6}
}

@misc{BloomOracle2026,
  author       = {{Bloom Energy}},
  title        = {Bloom Energy and Oracle Expand Strategic Partnership
                  to Deploy up to 2.8 GW to Accelerate AI Infrastructure
                  Build-Out},
  year         = {2026},
  month        = {April},
  url          = {https://www.bloomenergy.com/news/bloom-energy-and-oracle-expand-strategic-partnership-to-deploy-up-to-2-8-gw-to-accelerate-ai-infrastructure-build-out/}
}

@techreport{BCC2025SOFCMarket,
  author      = {{BCC Research}},
  title       = {Solid Oxide Fuel Cells: Technologies and Global Markets},
  institution = {BCC Research},
  year        = {2025},
  month       = {November},
  number      = {EGY048F},
  note        = {Global SOFC market projected to grow from USD 3.2 billion
                 in 2025 to USD 9.6 billion in 2030, at a 24.4\% CAGR},
  url         = {https://www.bccresearch.com/market-research/energy-and-resources/solid-oxide-fuel-cells-tech-markets-report.html}
}

@article{WeiEtAl1990SQS,
  author  = {Wei, S.-H. and Ferreira, L. G. and Bernard, James E.
             and Zunger, Alex},
  title   = {Electronic Properties of Random Alloys:
             Special Quasirandom Structures},
  journal = {Physical Review B},
  volume  = {42},
  number  = {15},
  pages   = {9622--9649},
  year    = {1990},
  doi     = {10.1103/PhysRevB.42.9622}
}

@article{HenkelmanEtAl2000,
  author  = {Henkelman, Graeme and Uberuaga, Blas P.
             and J{\'o}nsson, Hannes},
  title   = {A Climbing Image Nudged Elastic Band Method for Finding
             Saddle Points and Minimum Energy Paths},
  journal = {The Journal of Chemical Physics},
  volume  = {113},
  number  = {22},
  pages   = {9901--9904},
  year    = {2000},
  doi     = {10.1063/1.1329672}
}

@article{MarzariEtAl2012,
  author  = {Marzari, Nicola and Mostofi, Arash A. and Yates, Jonathan R.
             and Souza, Ivo and Vanderbilt, David},
  title   = {Maximally Localized Wannier Functions:
             Theory and Applications},
  journal = {Reviews of Modern Physics},
  volume  = {84},
  number  = {4},
  pages   = {1419--1475},
  year    = {2012},
  doi     = {10.1103/RevModPhys.84.1419}
}

@article{AichhornEtAl2016,
  author  = {Aichhorn, Markus and Pourovskii, Leonid and Seth, Priyanka
             and Vildosola, Veronica and Zingl, Manuel and Peil, Oleg E.
             and Deng, Xiaoyu and Mravlje, Jernej and Kraberger, Gernot J.
             and Martins, Cyril and Ferrero, Michel and Parcollet, Olivier},
  title   = {{TRIQS/DFTTools}: A {TRIQS} Application for
             Ab Initio Calculations of Correlated Materials},
  journal = {Computer Physics Communications},
  volume  = {204},
  pages   = {200--208},
  year    = {2016},
  doi     = {10.1016/j.cpc.2016.03.014}
}

@article{KarpEtAl2021,
  author  = {Karp, Jonathan and Hampel, Alexander and Millis, Andrew J.},
  title   = {Dependence of {DFT+DMFT} Results on the Construction
             of the Correlated Orbitals},
  journal = {Physical Review B},
  volume  = {103},
  number  = {19},
  pages   = {195101},
  year    = {2021},
  doi     = {10.1103/PhysRevB.103.195101}
}

@article{AryasetiawanEtAl2004,
  author  = {Aryasetiawan, F. and Imada, M. and Georges, A.
             and Kotliar, G. and Biermann, S. and Lichtenstein, A. I.},
  title   = {Frequency-Dependent Local Interactions and Low-Energy
             Effective Models from Electronic Structure Calculations},
  journal = {Physical Review B},
  volume  = {70},
  number  = {19},
  pages   = {195104},
  year    = {2004},
  doi     = {10.1103/PhysRevB.70.195104}
}

@article{Haule2015,
  author  = {Haule, Kristjan},
  title   = {Exact Double Counting in Combining the Dynamical Mean-Field
             Theory and the Density Functional Theory},
  journal = {Physical Review Letters},
  volume  = {115},
  number  = {19},
  pages   = {196403},
  year    = {2015},
  doi     = {10.1103/PhysRevLett.115.196403}
}

@article{GeorgesEtAl1996,
  author  = {Georges, Antoine and Kotliar, Gabriel and Krauth, Werner
             and Rozenberg, Marcelo J.},
  title   = {Dynamical Mean-Field Theory of Strongly Correlated Fermion
             Systems and the Limit of Infinite Dimensions},
  journal = {Reviews of Modern Physics},
  volume  = {68},
  number  = {1},
  pages   = {13--125},
  year    = {1996},
  doi     = {10.1103/RevModPhys.68.13}
}

@article{KotliarEtAl2006,
  author  = {Kotliar, Gabriel and Savrasov, Sergej Y. and Haule, Kristjan
             and Oudovenko, Viktor S. and Parcollet, Olivier
             and Marianetti, C. A.},
  title   = {Electronic Structure Calculations with Dynamical
             Mean-Field Theory},
  journal = {Reviews of Modern Physics},
  volume  = {78},
  number  = {3},
  pages   = {865--951},
  year    = {2006},
  doi     = {10.1103/RevModPhys.78.865}
}

@article{BauerEtAl2016,
  author  = {Bauer, Bela and Wecker, Dave and Millis, Andrew J.
             and Hastings, Matthew B. and Troyer, Matthias},
  title   = {Hybrid Quantum-Classical Approach to Correlated Materials},
  journal = {Physical Review X},
  volume  = {6},
  number  = {3},
  pages   = {031045},
  year    = {2016},
  doi     = {10.1103/PhysRevX.6.031045}
}

@article{Biermann2003GWDMFT,
  author  = {Biermann, S. and Aryasetiawan, F. and Georges, A.},
  title   = {First-Principles Approach to the Electronic Structure of Strongly Correlated Systems: Combining the {GW} Approximation and Dynamical Mean-Field Theory},
  journal = {Physical Review Letters},
  volume  = {90}, number = {8}, pages = {086402}, year = {2003},
  doi     = {10.1103/PhysRevLett.90.086402}
}

@article{Kananenka2015SEET,
  author  = {Kananenka, Alexei A. and Gull, Emanuel and Zgid, Dominika},
  title   = {Systematically improvable multiscale solver for correlated electron systems},
  journal = {Physical Review B},
  volume  = {91}, number = {12}, pages = {121111(R)}, year = {2015},
  doi     = {10.1103/PhysRevB.91.121111}
}

@article{Rusakov2019SEETPeriodic,
  author  = {Rusakov, Alexander A. and Iskakov, Sergei and Tran, Lan Nguyen and Zgid, Dominika},
  title   = {Self-Energy Embedding Theory ({SEET}) for Periodic Systems},
  journal = {Journal of Chemical Theory and Computation},
  volume  = {15}, number = {1}, pages = {229--240}, year = {2019},
  doi     = {10.1021/acs.jctc.8b00927}
}

@article{Iskakov2020SEETNiO,
  author  = {Iskakov, Sergei and Yeh, Chia-Nan and Gull, Emanuel and Zgid, Dominika},
  title   = {Ab initio self-energy embedding for the photoemission spectra of {NiO} and {MnO}},
  journal = {Physical Review B},
  volume  = {102}, number = {8}, pages = {085105}, year = {2020},
  doi     = {10.1103/PhysRevB.102.085105}
}

@article{Dang2014Covalency,
  author  = {Dang, Hung T. and Millis, Andrew J. and Marianetti, Chris A.},
  title   = {Covalency and the metal-insulator transition in titanate and vanadate perovskites},
  journal = {Physical Review B},
  volume  = {89}, number = {16}, pages = {161113(R)}, year = {2014},
  doi     = {10.1103/PhysRevB.89.161113}
}

@article{Hashimoto2010LSCF,
  author  = {Hashimoto, Shin-ichi and Fukuda, Yasuhiro and Kuhn, Melanie and Sato, Kazuhisa and Yashiro, Keiji and Mizusaki, Junichiro},
  title   = {Oxygen nonstoichiometry and thermo-chemical stability of {La$_{0.6}$Sr$_{0.4}$Co$_{1-y}$Fe$_y$O$_{3-\delta}$} (y = 0.2, 0.4, 0.6, 0.8)},
  journal = {Solid State Ionics},
  volume  = {181}, number = {37--38}, pages = {1713--1719}, year = {2010},
  doi     = {10.1016/j.ssi.2010.09.024}
}

@article{Mizusaki1989LSC,
  author  = {Mizusaki, Junichiro and Mima, Yasuo and Yamauchi, Shigeru and Fueki, Kazuo and Tagawa, Hiroaki},
  title   = {Nonstoichiometry of the perovskite-type oxides {La$_{1-x}$Sr$_x$CoO$_{3-\delta}$}},
  journal = {Journal of Solid State Chemistry},
  volume  = {80}, number = {1}, pages = {102--111}, year = {1989},
  doi     = {10.1016/0022-4596(89)90036-4}
}

@article{Gryaznov2012LSCF,
  author        = {Gryaznov, Denis and Finnis, Mike W. and Evarestov, Robert A. and Maier, Joachim},
  title         = {Oxygen vacancy formation energies in {Sr}-doped complex perovskites: ab initio thermodynamic study},
  journal       = {arXiv preprint arXiv:1206.6242},
  year          = {2012},
  eprint        = {1206.6242},
  archivePrefix = {arXiv},
  primaryClass  = {cond-mat.mtrl-sci}
}

@article{Wei2025LSCF,
  author        = {Wei, Wei and Fuchs, Florian and Zienert, Andreas and Hu, Xiao and Schuster, J{\"o}rg},
  title         = {First-principles analysis of the effect of magnetic states on the oxygen vacancy formation energy in doped {La$_{0.5}$Sr$_{0.5}$CoO$_3$} perovskite},
  journal       = {arXiv preprint arXiv:2507.07614},
  year          = {2025},
  eprint        = {2507.07614},
  archivePrefix = {arXiv},
  primaryClass  = {cond-mat.mtrl-sci}
}

@article{Chakrabarti2016LaCoO3,
  author  = {Chakrabarti, Bismayan and Birol, Turan and Haule, Kristjan},
  title   = {Role of entropy and structural parameters in the spin-state transition of {LaCoO$_3$}},
  journal = {Physical Review Materials},
  volume  = {1},
  number  = {6},
  pages   = {064403},
  year    = {2017},
  doi     = {10.1103/PhysRevMaterials.1.064403}
}

@article{Augustinsky2013LaCoO3,
  author  = {Augustinsk{\'y}, P. and K{\v{r}}{\'a}pek, V. and Kune{\v{s}}, J.},
  title   = {Doping induced spin state transition in {LaCoO$_3$}: dynamical mean-field study},
  journal = {Physical Review Letters},
  volume  = {110},
  number  = {26},
  pages   = {267204},
  year    = {2013},
  doi     = {10.1103/PhysRevLett.110.267204}
}

@article{WernerMillis2010DynamicalScreening,
  author  = {Werner, Philipp and Millis, Andrew J.},
  title   = {Dynamical screening in correlated electron materials},
  journal = {Physical Review Letters},
  volume  = {104},
  number  = {14},
  pages   = {146401},
  year    = {2010},
  doi     = {10.1103/PhysRevLett.104.146401}
}

@misc{ChenEtAl2023GibbsSampler,
  author        = {Chen, Chi-Fang and Kastoryano, Michael J. and Gily{\'e}n, Andr{\'a}s},
  title         = {An efficient and exact noncommutative quantum {Gibbs} sampler},
  year          = {2023},
  eprint        = {2311.09207},
  archivePrefix = {arXiv},
  primaryClass  = {quant-ph},
  url           = {https://arxiv.org/abs/2311.09207}
}

@article{Shinaoka2015SignProblem,
  author  = {Shinaoka, Hiroshi and Nomura, Yusuke and Biermann, Silke and Troyer, Matthias and Werner, Philipp},
  title   = {Negative sign problem in continuous-time quantum {Monte} {Carlo}: Optimal choice of single-particle basis for impurity problems},
  journal = {Physical Review B},
  volume  = {92},
  pages   = {195126},
  year    = {2015},
  doi     = {10.1103/PhysRevB.92.195126}
}

@article{Casula2012Bandwidth,
  author  = {Casula, M. and Werner, Ph. and Vaugier, L. and Aryasetiawan, F. and Miyake, T. and Millis, A. J. and Biermann, S.},
  title   = {Low-energy models for correlated materials: Bandwidth renormalization from {Coulombic} screening},
  journal = {Physical Review Letters},
  volume  = {109},
  pages   = {126408},
  year    = {2012},
  doi     = {10.1103/PhysRevLett.109.126408},
  eprint  = {1204.4900},
  archivePrefix = {arXiv}
}

@article{BuxbaumEtAl2025,
  author  = {Buxbaum, Barbara and Egger, Andreas
             and Kramer-Sch{\"o}ggl, Johanna
             and Falah Chamasemani, Fereshteh
             and Brunner, Roland and Bucher, Edith},
  title   = {Electrochemical performance and advanced
             3D reconstruction of {LSCF-GDC} air electrodes
             for solid oxide electrolysis cells},
  journal = {Journal of Power Sources},
  volume  = {657},
  pages   = {238174},
  year    = {2025},
  doi     = {10.1016/j.jpowsour.2025.238174},
  url     = {https://doi.org/10.1016/j.jpowsour.2025.238174}
}

@misc{ArunachalamEtAl2026,
  author        = {Arunachalam, Srinivasan and Bravyi, Sergey and Chowdhury, Anirban
                   and Dutt, Arkopal and Gheorghiu, Alexandru and Li, Zhi},
  title         = {Quantum impurity models: easy at equilibrium, universal in motion},
  year          = {2026},
  eprint        = {2610.02130},
  archivePrefix = {arXiv},
  primaryClass  = {quant-ph},
  doi           = {10.48550/arXiv.2610.02130},
  url           = {https://arxiv.org/abs/2610.02130}
}

@article{ZgidGull2017,
  author  = {Zgid, Dominika and Gull, Emanuel},
  title   = {Finite temperature quantum embedding theories for correlated systems},
  journal = {New Journal of Physics},
  volume  = {19},
  number  = {2},
  pages   = {023047},
  year    = {2017},
  doi     = {10.1088/1367-2630/aa5d34},
  eprint  = {1611.00395},
  archivePrefix = {arXiv}
}

@article{Almbladh1999Psi,
  author  = {Almbladh, C.-O. and von Barth, U. and van Leeuwen, R.},
  title   = {Variational total energies from {$\Phi$}- and {$\Psi$}-derivable theories},
  journal = {International Journal of Modern Physics B},
  volume  = {13},
  number  = {5n06},
  pages   = {535--541},
  year    = {1999},
  doi     = {10.1142/S0217979299000436}
}

@article{PhysRevB.92.115123,
  title = {Double-expansion impurity solver for multiorbital models with dynamically screened $U$ and $J$},
  author = {Steiner, Karim and Nomura, Yusuke and Werner, Philipp},
  journal = {Phys. Rev. B},
  volume = {92},
  issue = {11},
  pages = {115123},
  numpages = {15},
  year = {2015},
  month = {Sep},
  publisher = {American Physical Society},
  doi = {10.1103/PhysRevB.92.115123},
  url = {https://link.aps.org/doi/10.1103/PhysRevB.92.115123}
}

@article{Eidelstein2020Inchworm,
  author  = {Eidelstein, Eitan and Gull, Emanuel and Cohen, Guy},
  title   = {Multiorbital Quantum Impurity Solver for General Interactions and Hybridizations},
  journal = {Physical Review Letters},
  volume  = {124},
  pages   = {206405},
  year    = {2020},
  doi     = {10.1103/PhysRevLett.124.206405},
  eprint  = {1907.08570},
  archivePrefix = {arXiv}
}

@article{Strand2024InchwormQMC,
  author  = {Strand, Hugo U. R. and Kleinhenz, Joseph and Krivenko, Igor},
  title   = {Inchworm quasi Monte Carlo for quantum impurities},
  journal = {Physical Review B},
  volume  = {110},
  pages   = {L121120},
  year    = {2024},
  eprint  = {2310.16957},
  archivePrefix = {arXiv}
}

@article{Werner2012Satellites,
  author  = {Werner, Philipp and Casula, Michele and Miyake, Takashi and Aryasetiawan, Ferdi and Millis, Andrew J. and Biermann, Silke},
  title   = {Satellites and large doping- and temperature-dependence of electronic properties in hole-doped {BaFe$_2$As$_2$}},
  journal = {Nature Physics},
  volume  = {8},
  pages   = {331--337},
  year    = {2012},
  eprint  = {1107.3128},
  archivePrefix = {arXiv}
}

@article{WangMaxischCeder2006,
  author  = {Wang, L. and Maxisch, T. and Ceder, G.},
  title   = {Oxidation energies of transition metal oxides within the {GGA}+{U} framework},
  journal = {Physical Review B},
  volume  = {73},
  pages   = {195107},
  year    = {2006},
  doi     = {10.1103/PhysRevB.73.195107}
}

\newpage 
\appendix

\section*{Appendix}
\addcontentsline{toc}{section}{Appendix}

\setcounter{secnumdepth}{2}
\renewcommand{\thesubsection}{\Alph{subsection}}

\subsection{Crystallographic and atomistic construction details}
\label{app:crystallography}

\subsubsection{Generic procedure}
The first input is a pair of experimentally relevant crystal structures, for the composition and operating conditions of interest. A crystal is described by a unit cell that contains an atomic basis and is repeated periodically \cite{InternationalTables2016}.

\begin{figure}[H]
    \centering
    \includegraphics[width=\textwidth]{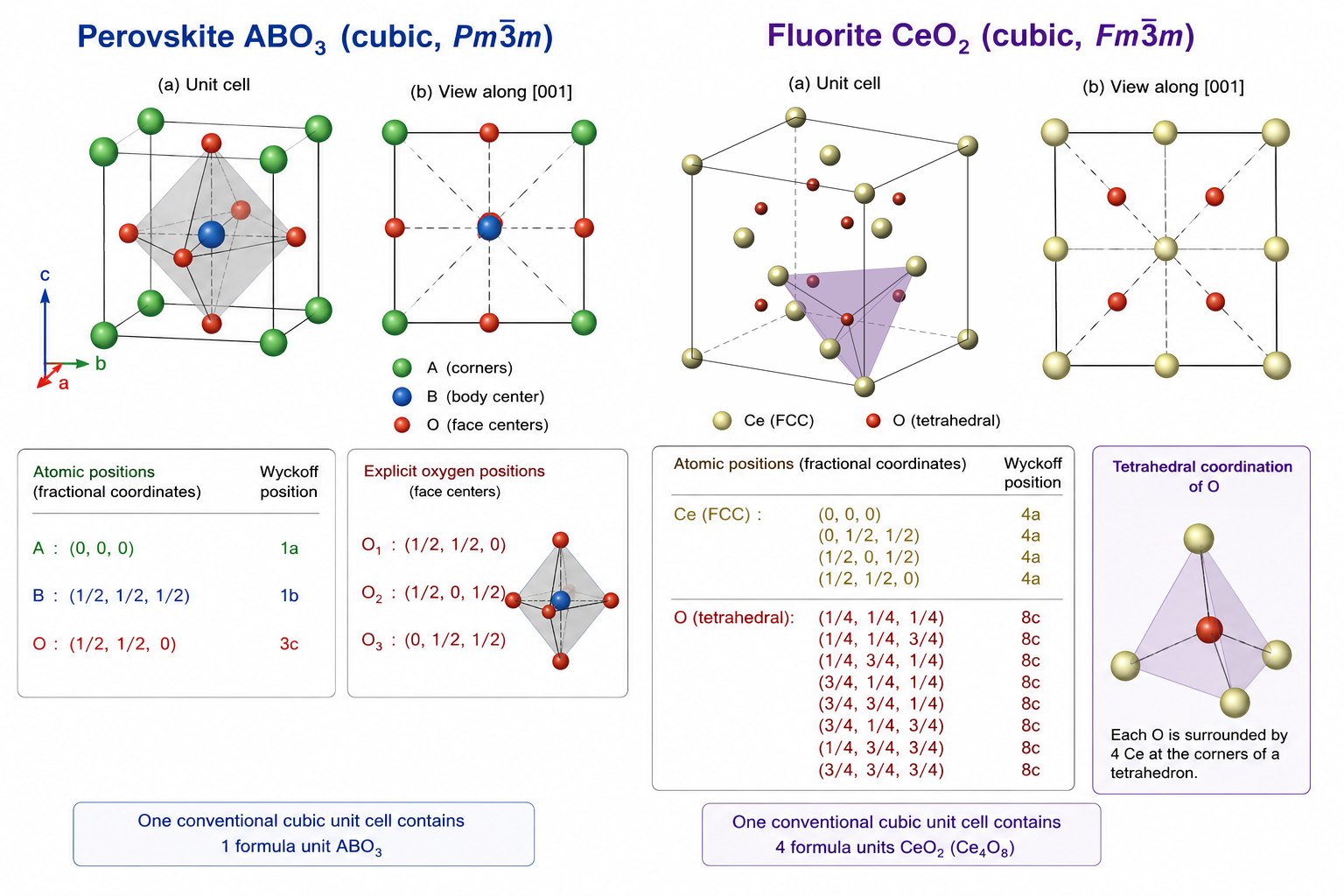}
    \caption{
    Ideal bulk crystal structures used as the starting point for the atomistic models. 
    Left: cubic perovskite \(ABO_3\) (\(Pm\bar{3}m\)), with \(A\) cations at the cube corners, the \(B\) cation at the body center, and oxygen atoms at the face centers. 
    Right: cubic fluorite \(\mathrm{CeO_2}\) (\(Fm\bar{3}m\)), with Ce cations forming a face-centered-cubic sublattice and oxygen atoms occupying the tetrahedral sites. 
    The actual LSCF and GDC structures are obtained from these parent lattices by introducing the appropriate cation substitutions and oxygen non-stoichiometry, followed by structural relaxation.}
    \label{fig:bulk_geometries}
\end{figure}

For an ideal cubic perovskite \(ABO_3\), where $A$ and $B$ denote different chemical elements whereas $O$ always denotes oxygen atoms, Fig.~\ref{fig:bulk_geometries} left, the conventional fractional coordinates are
\[
A:(0,0,0),
\qquad
B:\left(\frac{1}{2},\frac{1}{2},\frac{1}{2}\right),
\]
\[
O_1:\left(\frac{1}{2},\frac{1}{2},0\right),
\quad
O_2:\left(\frac{1}{2},0,\frac{1}{2}\right),
\quad
O_3:\left(0,\frac{1}{2},\frac{1}{2}\right).
\]
The corner \(A\) atoms and face-centered oxygen atoms are shared between neighboring cells, so the basis contains one \(ABO_3\) formula unit. 
A \(2\times2\times1\) supercell therefore repeats this atomic basis four times and contains four \(ABO_3\) formula units.

Starting from the ideal perovskite \(ABO_3\), the LSCF composition is obtained by substituting the \(A\)- and \(B\)-site cations:
\[
\mathrm{La}_{0.6}\mathrm{Sr}_{0.4} \mathrm{Co}_{0.2}\mathrm{Fe}_{0.8} \mathrm{O}_{3-\delta}.
\]

On the \(A\)-site, \(\mathrm{Sr}^{2+}\) substitutes for \(\mathrm{La}^{3+}\). Because \(\mathrm{Sr}^{2+}\) carries one positive charge less than \(\mathrm{La}^{3+}\), this aliovalent substitution must be compensated by changes elsewhere in the crystal. 
Charge compensation occurs through a combination of changes in the oxidation states of the Co/Fe \(B\)-site cations and the formation of oxygen vacancies.
The latter produces the oxygen non-stoichiometry represented by \(\delta\) in \(O_{3-\delta}\).
 
On the \(B\)-site, Co and Fe occupy the transition-metal sites in the nominal ratio. 
For example, a supercell containing \(20\) perovskite formula units can represent the nominal composition explicitly as
\[
\mathrm{La}_{12} \mathrm{Sr}_{8} \mathrm{Co}_{4} \mathrm{Fe}_{16} \mathrm{O}_{60-N_{\mathrm{vac}}},
\]
where \(N_{\mathrm{vac}}\) is the number of oxygen atoms removed from the oxygen sublattice. 
The corresponding oxygen non-stoichiometry is
\begin{equation}
    \delta=\frac{N_{\mathrm{vac}}}{20}.
\end{equation}
Removing one oxygen atom gives \(\delta=0.05,\) corresponding to
\[
\mathrm{La}_{0.6}\mathrm{Sr}_{0.4} \mathrm{Co}_{0.2}\mathrm{Fe}_{0.8} \mathrm{O}_{2.95}.
\]

The particular La/Sr and Co/Fe site occupations and the locations of the oxygen vacancies must be specified explicitly in the atomistic supercell and the resulting structure subsequently relaxed with DFT.

For an ideal cubic fluorite electrolyte \(M O_2\), such as ceria \(\mathrm{CeO_2}\), Fig.~\ref{fig:bulk_geometries} right, the conventional fractional coordinates of the cations are

\[
M_1:(0,0,0),
\qquad
M_2:\left(0,\frac{1}{2},\frac{1}{2}\right),
\]
\[
M_3:\left(\frac{1}{2},0,\frac{1}{2}\right),
\qquad
M_4:\left(\frac{1}{2},\frac{1}{2},0\right),
\]
while the oxygen atoms occupy the tetrahedral sites
\[
O_1:\left(\frac{1}{4},\frac{1}{4},\frac{1}{4}\right),
\quad
O_2:\left(\frac{1}{4},\frac{1}{4},\frac{3}{4}\right),
\]
\[
O_3:\left(\frac{1}{4},\frac{3}{4},\frac{1}{4}\right),
\quad
O_4:\left(\frac{3}{4},\frac{1}{4},\frac{1}{4}\right),
\]
\[
O_5:\left(\frac{3}{4},\frac{3}{4},\frac{1}{4}\right),
\quad
O_6:\left(\frac{3}{4},\frac{1}{4},\frac{3}{4}\right),
\]
\[
O_7:\left(\frac{1}{4},\frac{3}{4},\frac{3}{4}\right),
\quad
O_8:\left(\frac{3}{4},\frac{3}{4},\frac{3}{4}\right).
\]

The cations form a face-centered-cubic sublattice, while the oxygen atoms occupy the eight tetrahedral sites of the conventional cell. 
One conventional fluorite cell therefore contains four \(MO_2\) formula units, i.e.~\(M_4O_8\).

For the electrolyte GDC, the parent fluorite structure is modified by replacing a fraction of the \(\mathrm{Ce}^{4+}\) cations by \(\mathrm{Gd}^{3+}\). 
For example,
\[
\mathrm{Ce}_{0.9}\mathrm{Gd}_{0.1}\mathrm{O}_{1.95}
\]
corresponds to replacing \(10\%\) of the Ce sites by Gd. 
Charge neutrality requires approximately one oxygen vacancy for every two \(\mathrm{Gd}^{3+}\) ions substituted for \(\mathrm{Ce}^{4+}\):
\[
2\,\mathrm{Gd}_{\mathrm{Ce}} \quad\longrightarrow\quad V_{\mathrm O},
\]
in standard Kröger–Vink notation.
 
Because a single conventional fluorite cell contains only four cation sites, a larger supercell is required to represent the \(10\%\) Gd concentration explicitly. 
For example, a supercell containing \(20\) cation sites can be constructed from the ideal fluorite lattice and populated with
\[
18\,\mathrm{Ce}, \qquad 2\,\mathrm{Gd}, \qquad 39\,\mathrm{O}.
\]
Starting from the ideal fluorite coordinates, two Ce atoms are replaced by Gd and one oxygen atom is removed from one of the tetrahedral oxygen sites. 
The particular locations of the Gd dopants and oxygen vacancy constitute additional atomistic degrees of freedom and should be sampled over several distinct configurations. 
The resulting structure is subsequently relaxed with DFT, so that the final atomic positions generally differ from the ideal fluorite coordinates given above.

\begin{figure}[H]
    \centering
    \includegraphics[width=\textwidth]{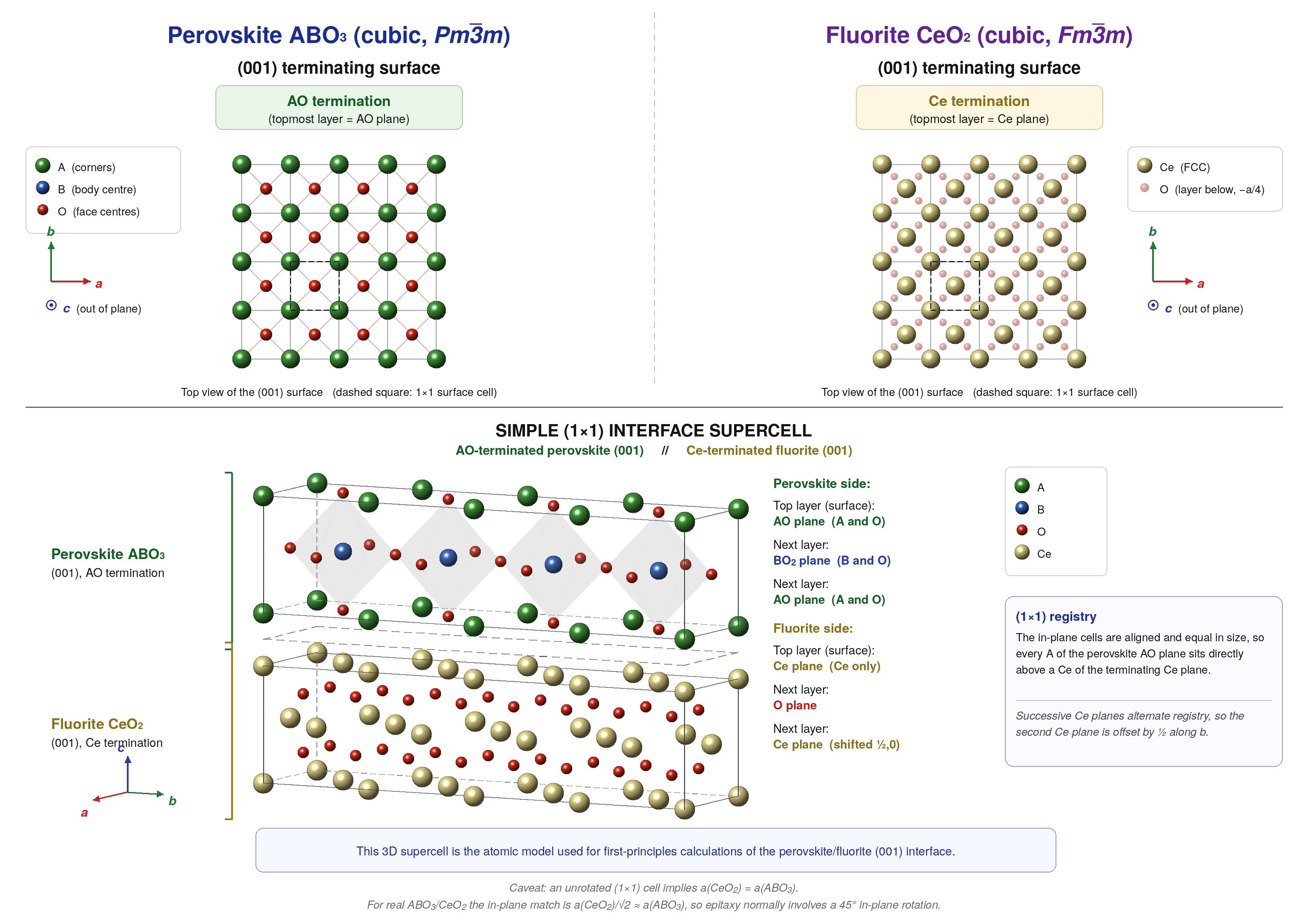}
    \caption{
    Illustration of the construction of an electrode/electrolyte interface supercell from the parent perovskite and fluorite structures. 
    The example shows a (001)-oriented, \(AO\)-terminated perovskite interfaced with a Ce-terminated (001) fluorite structure. 
    In an application to LSCF/GDC, the in-plane surface cells are first made commensurate through an appropriate choice of supercell, relative rotation, and small lattice strain. 
    The terminating planes and their lateral registry are then chosen, an initial interface separation is specified, and the complete atomic structure is relaxed with DFT. 
    The resulting relaxed supercell provides the atomistic input for the subsequent electronic-structure and embedding calculations.}
    \label{fig:interface_geometry}
\end{figure}

Building an interface from these two bulk structures requires four choices, illustrated in Fig.~\ref{fig:interface_geometry}.

\paragraph{Orientation.}
The orientation specifies which crystallographic plane of the electrode faces which plane of the electrolyte, for example
\[
(001)_{\mathrm{electrode}} \parallel (001)_{\mathrm{electrolyte}}.
\]
 
\paragraph{Termination.}
The termination specifies the last atomic plane of each crystal. Along the \([001]\) direction, an ideal perovskite consists of alternating planes
\[
AO \;|\; BO_2 \;|\; AO \;|\; BO_2 \;|\cdots.
\]
It can therefore terminate in either an \(AO\) or a \(BO_2\) plane.
 
\paragraph{Registry.}
The registry specifies the lateral translation between the two terminated surfaces.
A transition-metal ion at the interface may, for example, face an electrolyte oxygen, an electrolyte cation, or a hollow site. 
These registries generally have different energies.
 
\paragraph{Commensurate surface cell.}
The two surface lattices must be represented in a common periodic cell.
Integer repetitions and relative rotations of the two surface cells are searched until their in-plane vectors match within an acceptable strain \cite{ZurMcGill1984}. 
If the surface matrices of the two materials, whose columns are the in-plane lattice vectors, are \(\mathbf{S}_{1}\) and \(\mathbf{S}_{2}\), the search seeks integer matrices \(\mathbf{M}_{1}\) and \(\mathbf{M}_{2}\) and a rotation \(R(\theta)\) such that
\begin{equation}
    \mathbf{S}_{1}\mathbf{M}_{1} \approx R(\theta)\mathbf{S}_{2}\mathbf{M}_{2}.
\end{equation}

For each plausible orientation, the workflow is:
\begin{enumerate}
    \item enumerate distinct terminations of both materials;
    \item identify low-strain commensurate surface cells;
    \item enumerate distinct lateral registries;
    \item relax each candidate interface using DFT;
    \item retain the lowest-energy structures and any near-degenerate
    alternatives.
\end{enumerate}

For a symmetric periodic supercell containing two equivalent interfaces, an interface-energy comparison can be written schematically as
\begin{equation}
    \gamma_{\mathrm{int}} = \frac{E_{\mathrm{cell}} - \sum_s N_s \mu_s }{2S},
\end{equation}
where \(S\) is the area of one interface, \(N_s\) is the number of atoms of species \(s\), and \(\mu_s\) is its chemical potential. 
If two candidate terminations contain different numbers of oxygen atoms, their stability must be compared at the oxygen chemical potential corresponding to the relevant temperature and oxygen partial pressure.

\subsubsection{Running example: an LSCF/GDC interface}
We use as a running example
\[
\underbrace{\mathrm{La}_{0.6}\mathrm{Sr}_{0.4} \mathrm{Co}_{0.2}\mathrm{Fe}_{0.8} \mathrm{O}_{3-\delta}}_{\mathrm{LSCF\,oxygen\,electrode}}
\;\Big|\;
\underbrace{\mathrm{Ce}_{0.9}\mathrm{Gd}_{0.1}\mathrm{O}_{1.95}}_{\mathrm{GDC\ electrolyte}}.
\]

LSCF/GDC interfaces are very relevant to solid-oxide cells and we have elaborated on the motivations underlying this choice in Section ~\ref{sec:motivation}. 

For the present study, we adopt an experimentally motivated epitaxial relationship in which the interface plane is the (001) plane of both materials:
\[
(001)_{\mathrm{pc}}^{\mathrm{LSCF}} \parallel (001)_{\mathrm{c}}^{\mathrm{GDC}},
\]
where ``pc'' denotes the pseudocubic representation of LSCF and ``c'' denotes the cubic fluorite cell of GDC \cite{LeeEtAl2016LSCFHeterostructure}.

The two crystals are rotated relative to one another within the interface plane. 
We choose the in-plane alignment
\[
[100]_{\mathrm{pc}}^{\mathrm{LSCF}} \parallel [110]_{\mathrm{c}}^{\mathrm{GDC}},
\]
and
\[
[010]_{\mathrm{pc}}^{\mathrm{LSCF}} \parallel [\bar{1}10]_{\mathrm{c}}^{\mathrm{GDC}}.
\]
Thus, the LSCF and GDC (001) surfaces remain parallel, while their in-plane crystallographic axes are rotated by \(45^\circ\).

This rotation provides a favorable lattice match. Representative lattice parameters are approximately \cite{HardyEtAl2012LSCF,LeeEtAl2017GDC}
\begin{equation}
    a_{\mathrm{LSCF,pc}} \simeq 3.925\ \text{\AA}, 
    \qquad
    a_{\mathrm{GDC,c}} \simeq 5.417\ \text{\AA}.
\end{equation}

Under the \(45^\circ\) in-plane rotation, the relevant GDC periodicity is
\begin{equation}
    \frac{a_{\mathrm{GDC,c}}}{\sqrt{2}} \simeq 3.83\ \text{\AA},
\end{equation}
which can be compared directly with the LSCF pseudocubic lattice parameter.
The resulting lattice mismatch is approximately
\begin{equation}
    \epsilon_{\parallel} = \frac{a_{\mathrm{LSCF,pc}} - a_{\mathrm{GDC,c}}/\sqrt{2}}{a_{\mathrm{GDC,c}}/\sqrt{2}} \simeq 2.5\%.
\end{equation}

A commensurate interface supercell can therefore be constructed by repeating the LSCF surface cell twice along each in-plane direction, to ensure that it contains at least one full cell for each crystal. The supercell's in-plane dimensions are approximately
\begin{equation}
    2a_{\mathrm{LSCF,pc}} \simeq 7.85\ \text{\AA}.
\end{equation}
The corresponding GDC surface cell spans approximately
\begin{equation}
    \sqrt{2}a_{\mathrm{GDC,c}} \simeq 7.66\ \text{\AA}
\end{equation}
along the matched directions. 
A common in-plane lattice parameter is then imposed, with the remaining small mismatch accommodated by straining one material.

The out-of-plane thickness is chosen independently. For example, if the LSCF slab contains five pseudocubic cells along the interface normal, the \(2\times2\times5\) LSCF block contains 20 perovskite formula units. 
Before introducing oxygen non-stoichiometry, the nominal LSCF composition can therefore be represented explicitly as
\[
12\ \mathrm{La},
\qquad
8\ \mathrm{Sr},
\qquad
4\ \mathrm{Co},
\qquad
16\ \mathrm{Fe},
\qquad
60\ \mathrm{O}.
\]

The GDC block is constructed from the fluorite lattice with the same commensurate in-plane periodicity and sufficient thickness to recover bulk-like behavior away from the interface. 
To represent \(\mathrm{Ce}_{0.9}\mathrm{Gd}_{0.1}\mathrm{O}_{1.95}\), the number of cation sites should be compatible with the \(10\%\) Gd concentration. 
For example, a block containing 20 cation sites can contain
\[
18\ \mathrm{Ce},
\qquad
2\ \mathrm{Gd},
\qquad
39\ \mathrm{O},
\]
where the missing oxygen provides the charge-compensating oxygen vacancy associated with the two \(\mathrm{Gd}^{3+}\) substitutions.

The orientation relationship is held fixed in the baseline calculation.
Within this orientation, several atomistic degrees of freedom must still be explored. 
On the LSCF side, the (001) surface can terminate in either an
\[
(\mathrm{La},\mathrm{Sr})O
\]
plane or a
\[
(\mathrm{Co},\mathrm{Fe})O_2
\]
plane.
Corresponding chemically plausible GDC (001) terminations should be constructed, including different interfacial oxygen contents where relevant. For each pair of terminating planes, several lateral registries are generated by translating one surface relative to the other within the interface plane.

The La/Sr, Co/Fe, and Ce/Gd occupations must also be assigned to explicit atomic sites. 
A pilot calculation may use a low-energy ordered configuration, while production calculations should sample several configurations or use a special quasirandom structure to represent substitutional disorder \cite{WeiEtAl1990SQS}. 
Likewise, distinct oxygen-vacancy locations should be considered because their stability depends on the local cation environment and distance from the interface.
 
Each candidate interface is constructed with a physically reasonable initial separation between the two terminating atomic planes and subsequently relaxed with DFT. 
The interface separation, local bond lengths, octahedral distortions, and atomic reconstruction are therefore outputs of the structural relaxation rather than fixed parameters of the final model. 
The relaxed candidate structures are compared using their interface energies, and the lowest-energy structures, together with any relevant near-degenerate alternatives, are retained for the subsequent embedding calculations.
 
A symmetric periodic supercell can be represented schematically as
\[
\begin{array}{c}
\text{bulk-like LSCF layers}
\\
\hline
\text{interfacial LSCF layer}
\\
\text{interfacial GDC layer}
\\
\hline
\text{bulk-like GDC layers}.
\end{array}
\]
The LSCF and GDC regions are made sufficiently thick that the layers farthest from the interface recover their respective bulk-like structural and electronic properties.
 
Thus, the baseline atomistic construction is fixed to the experimentally motivated orientation relationship while termination, lateral registry, cation ordering, oxygen-vacancy configuration, and the relaxed interface separation constitute the principal atomistic variables explored by the DFT workflow.

\subsection{Selection of vacancy and migration geometries}
\label{app:select-vacancy}
The calculation should not assume that one arbitrarily selected oxygen atom is representative. 
Oxygen sites are first divided into chemically distinct
classes according to:
\begin{itemize}
    \item distance from the LSCF/GDC interface;
    \item identities of neighboring transition-metal cations;
    \item local La/Sr or Ce/Gd environment;
    \item proximity to an existing dopant-compensating vacancy;
    \item local bond lengths and octahedral distortions.
\end{itemize}
 
In LSCF, an oxygen atom typically bridges two \(B\)-site cations. 
Distinct
local environments therefore include
\[
\mathrm{Fe{-}O{-}Fe},
\qquad
\mathrm{Fe{-}O{-}Co},
\qquad
\mathrm{Co{-}O{-}Co}.
\]
These environments are further distinguished by their distance from the interface and their nearby La/Sr arrangement.
 
The practical procedure is:
\begin{enumerate}
    \item enumerate all symmetry- and chemistry-distinct oxygen sites;
    \item remove one oxygen from each site and relax the resulting structure
    at the DFT level;
    \item rank the sites by their DFT vacancy-formation energies;
    \item apply the Tier T1 embedding to the lowest-energy
    sites and to any sites directly relevant to transport across the interface.
\end{enumerate}
 
For the running example, consider an oxygen atom \(O^{*}\)
in the LSCF \(B\mathrm{O}_2\) layer closest to GDC, with local coordination
\[
\mathrm{Fe}_{1} - O^{*} - \mathrm{Fe}_{2}.
\]
The two principal structures for vacancy formation are then:
\begin{equation}
    g_{\mathrm{pristine}} = \text{relaxed LSCF/GDC interface containing }O^{*},
\end{equation}
and
\begin{equation}
    g_{\mathrm{vac}} = \text{the same interface after removing }O^{*}.
\end{equation}
 
The vacancy-formation energy is obtained from the converged correlated energies of these two structures through Eq.~\eqref{eq:evac}. 
The selection of an Fe--O--Fe site is only an illustrative pilot choice; the preceding site enumeration determines whether it is in fact one of the relevant low-energy vacancies.
 
For migration, the initial state is a relaxed structure containing an oxygen vacancy at a particular lattice site. 
Consider, for example, two neighboring oxygen sites in the ideal cubic perovskite,
\[
O_1=\left(\frac{1}{2},\frac{1}{2},0\right),
\qquad
O_2=\left(\frac{1}{2},0,\frac{1}{2}\right).
\]
Suppose that \(O_1\) is occupied and \(O_2\) is vacant. 
The migration event corresponds to the oxygen atom moving from \(O_1\) into the vacancy at \(O_2\):
\[
\left(\frac{1}{2},\frac{1}{2},0\right)
\longrightarrow
\left(\frac{1}{2},0,\frac{1}{2}\right).
\]
After the hop, \(O_2\) is occupied and \(O_1\) has become the new vacancy site.
 
The transition state is not another stable lattice configuration and its coordinates are not specified in advance. 
During migration, the oxygen leaves its equilibrium lattice site and passes through an interstitial-like region between \(O_1\) and \(O_2\), locally distorting the surrounding crystal. 
The geometric midpoint,
\[
\left(\frac{1}{2},\frac{1}{4},\frac{1}{4}\right),
\]
can be used as an initial guess for an intermediate configuration, but the actual migration path generally deviates from this straight-line trajectory as the migrating oxygen and surrounding atoms relax.
 
A NEB calculation is therefore performed between the known initial and final geometries. 
Several intermediate atomic configurations are introduced and optimized collectively to obtain the minimum-energy migration path \cite{HenkelmanEtAl2000}. 
The highest-energy configuration along this optimized path is the transition-state geometry \(g_{\mathrm{TS}}\). 
The embedding is then applied to the relaxed initial geometry \(g_{\mathrm{IS}}\) and the NEB-derived transition-state geometry \(g_{\mathrm{TS}}\), giving the migration barrier
\begin{equation}
    E_{\mathrm{mig}}^\mathcal{I} = E_{\mathrm{T1}}(g_{\mathrm{TS}}) - E_{\mathrm{T1}}(g_{\mathrm{IS}}).
\end{equation}

The resulting structural dataset is therefore
\begin{equation}
\mathcal{G} = \left\{g_{\mathrm{pristine}}, g_{\mathrm{vac},1}, g_{\mathrm{vac},2}, \ldots, g_{\mathrm{IS},1}, g_{\mathrm{TS},1}, \ldots \right\}.
\end{equation}

\end{document}